\documentclass[graybox]{svmult}

\usepackage{mathptmx}       
\usepackage{helvet}         
\usepackage{courier}        
\usepackage{type1cm}        
\usepackage{makeidx}         
\usepackage{graphicx}        
\usepackage{multicol}        
\usepackage[bottom]{footmisc}
\usepackage{epstopdf}

\makeindex             

\begin{document}

\title*{Statistical methods in cosmology}
\author{Licia Verde}
\institute{Licia Verde \at ICREA \& ICE (IEEC-CSIC) and ICC UB, \email{verde@icc.ub.edu}}
%
%
\maketitle

\abstract{ The advent of large data-set in cosmology has meant that in the past 10 or 20  years our knowledge  and understanding of the Universe has changed not only quantitatively but also, and most importantly, qualitatively. Cosmologists are interested in studying the origin and evolution of the physical Universe. They rely on data where a host of useful  information is enclosed, but is encoded in a  non-trivial way. 
The challenges in extracting this information    must be overcome to  make the most of the  large experimental effort.
Even after having analyzed a decade or more of data and having  converged to a standard cosmological model (the so-called and highly successful LCDM model) we should keep in mind that this model is   described by 10 or more physical parameters and if we want to study deviations from the standard model the number of parameters is even larger. Dealing with  such a high dimensional parameter space and finding parameters constraints  is a challenge on itself.  In addition, as gathering data is such an expensive and difficult process,  cosmologists  want to be able to compare and combine different data sets both for testing for possible disagreements (which could indicate new physics)  and for improving  parameter determinations.  Finally, always because experiments are so expansive, cosmologists in many cases want to find out {\it a priori}, before actually doing the experiment, how much one would be able to learn from it.
For all these reasons,  more and more sophisiticated statistical techniques are being employed in cosmology, and it has become crucial  to know some statistical background to understand recent literature in the field.
Here,   I will introduce some statistical tools that any cosmologist should know 
about in order to be able to understand  recently published results from the 
analysis of  cosmological data sets. I will not present a complete and rigorous 
introduction to statistics as there are several good books which are reported in the references. The reader should refer to those. 
I will take a practical approach and  I will touch upon useful tools such as  statistical inference,  Bayesians vs Frequentist approach, chisquare and goodness of fit, confidence regions, likelihood, Fisher matrix approach,   Monte Carlo methods and  a brief introduction to model testing. Throughout, I will use practical examples often taken from recent literature to illustrate the use of such tools. Of course this will not be an exhaustive guide: it should be interpreted as a ``starting kit", and the reader is warmly encouraged to read the references to find out more.}

\section{Introduction}
\label{sec:1}

As cosmology has made the transition from a data-starved science to a data-driven science, the use of increasingly sophisticated statistical tools  has increased. 
As  explained in detail below, cosmology is intrinsically related to statistics, as theories of the origin and evolution of the Universe  do not predict, for example, that a particular galaxy will form at a specific point in space and time or that  a specific patch of the cosmic microwave background will have a given temperature: any theory will predict average statistical properties of our Universe, and we can only observe a particular realization of that.

It is often said that cosmology has entered the precision era:  ``precision"  requires a good knowledge of  the error-bars  and thus confidence intervals  of a measurement. This is an inherently statistical statement. We should try however to go even further, and achieve also ``accuracy" (although cosmology    does not have a particularly stellar track record in this regard). This requires quantifying systematic errors (beyond the statistical ones) and it is also requires statistical tools.
For all these reasons,  knowledge of basic statistical tools has become indispensable to understand the recent cosmological literature.

Examples of applications where  probability and statistics are crucial in Cosmology  are:  {\it i)} Is the universe homogenous and isotropic on large scales? {\it ii)} are the initial conditions consistent with being Gaussian? {\it iii)} is there a detection of non-zero tensor modes? {\it iv)} what is the value of the density parameter of the Universe $\Omega_m$ given the WMAP data for a LCDM model? {\it v)} what are the allowed values at a given confidence level for the primordial power spectrum spectral slope $n$?  {\it vi)}  what is the best fit  value of the dark energy equation of state parameter  $w$?  {\it vii)} Is a model with equation of state parameter different from -1 a better fit to the data than a model with non-zero curvature?  {\it viii)}  what will be the constraint on the parameter $w$ for a survey with given characteristic? 

The first three questions address the hypothesis testing issue. You have an hypothesis and you want to check wether the data are consistent with it. Sometimes, especially for addressing issues of ``detection"  you can  test the null hypothesis: assume the quantity is zero and test wether the data are consistent with it. 

The next three questions  are ``parameter estimation" problems: we have a model, in this example the LCDM model,  which is characterized by some free parameters  which we would like to measure. 

The next question, {\it vii)}, belong to ``model testing": we have two models and ask which one is a better fit to the data. Model testing comes in several different flavors: the two models to be considered may have different number of parameters or equal number of parameters, may be have some parameters in common or not etc.

Finally question {\it viii)} is on ``forecasting", which is particularly  useful for
or quickly forecasting the performance of future experiments and for  experimental design.

Here we will mostly concentrate in the issue of parameter estimation but also touch upon the other applications.

\section{Bayesian vs Frequentists}
\label{sec:2}
The world is divided in Frequentists\index{Frequentist} and Bayesians\index{Bayesian}. 
For Frequentsists probabilities ${\cal P}$  are frequencies of occurence:
\begin{equation}
{\cal P}=\frac{n}{N}
\end{equation}
where $n$ denotes the number of successes and $N$ the total number of trials.
Frequentists define probability\index{probability}  as the limit for the number of independent trials going to infinity.
Bayesians interpret probabilities as {\it degree of belief in a Hypothesis}. 

Let us say that $x$ is our random variable (event). Depending on the application, $x$ can be the number of photons hitting a detector, the matter density in a volume, the Cosmic Microwave Background temperature in a direction in the sky, etc.  The probability that $x$ takes a specific value is ${\cal P}(x)$ where ${\cal P}$ is called probability distribution. Note that probabilities  (the possible values of $x$) can be discrete or continuous. ${\cal P}(x)$ is a {\it probability density}: ${\cal P}(x) dx$ is the probability that the random variable $x$ takes a value between $x$ and $x+dx$.
Frequentists only consider probability distributions of events while Bayesians consider hypothesis as events.

For both, the rules of probability apply.
\begin{itemize}
\item{1.} ${\cal P}(x)\ge 0$
\item{2.} $\int _{-\infty}^{\infty}dx {\cal P}(x)=1$. In the discrete case $\int \longrightarrow \sum$.
\item{3.} For mutually exclusive events ${\cal P}(x_1 U x_2)\equiv {\cal P}(x_1.OR.x_2)={\cal P}(x_1)+{\cal P}(x_2)$
\item{4.} In general ${\cal P}(x_1,x_2)={\cal P}(x_1){\cal P}(x_1|x_2)$. In words: the probability of $x_1$ AND $x_2$ to happen is the probability of $x_1$ times the {\it conditional probability} of $x_2$ given that $x_1$ has already happened.
\end{itemize}
The last item deserves some discussion.  
Example here.
Only for independent events where ${\cal P}(x_2|x_1)={\cal P}(x_2)$ one can write ${\cal P}(x_1,x_2)={\cal P}(x_1){\cal P}(x_2)$.
Of course  in general one can always rewrite ${\cal P}(x_1,x_2)={\cal P}(x_1){\cal P}(x_1|x_2)$ by switching $x_1$and $x_2$.
If then one makes the apparently tautological identification that  ${\cal P}(x_1,x_2)= {\cal P}(x_2,x_1)$ and substitute $x_1\longrightarrow  D$ standing for {\it data} and $x_2\longrightarrow H$ standing for {\it hypothesis}, one gets {\bf Bayes theorem}\index{Bayes theorem} :
\begin{equation}
\label{Eq:bayesth}
{\cal P}(H|D)=\frac{{\cal P}(H) {\cal P}(D|H)}{{\cal P}(D)}
\end{equation}
${\cal P}(H|D)$ is called the {\bf posterior}\index{posterior}, ${\cal P}(D|H)$ is the {\bf likelihood}\index{likelihood} (the probability of the data given the hypothesis, and ${\cal P}(H)$ is called the {\bf prior}\index{prior}.
Note that here  explicitly we have probability and probability distribution of a hypothesis. 

\section{Bayesian approach and statistical inference}
Despite its simplicity, Bayes theorem is at the base of statistical inference.
For the Bayesian\index{Bayesian} point of view let us  use $D$ to indicate our data (or data set). The hypothesis $H$ can be a model, say for example the LCDM model, which is characterized by a set of parameters $\vec{\theta}$.  In the Bayesian framework what we want to know is ``what is the probability distribution for the model parameters given the data?" i.e. ${\cal P}(\vec{\theta}|D)$. From this information then we can extract the most likely value for the parameters and their confidence limits\footnote{At this point many Frequentists stop reading this document...}. However what we can compute accurately,  in most instances, is the likelihood,  which is related to the posterior by the prior. (At this point one assumes one has collected the data and so ${\cal P}(D)=1$). The prior however can be somewhat arbitrary. This is a crucial point to which we will return below. For now let us consider an example: the constraint from WMAP data on the integrated optical depth to the last scattering surface $\tau$. One could do the analysis using the variable $\tau$ itself, however one could also note that the temperature data (the angular power spectrum of the temperature fluctuations) on large scale depend approximately linearly on the variable $Z=\exp(-2 \tau)$. A third person would note that the polarization  data (in particular the EE angular power spectrum) depends roughly linearly on $\tau^2$. So person one  could use a uniform prior in $\tau$, person two a uniform prior in $\exp(-2 \tau)$ and person three in $\tau^2$. What is the relation between ${\cal P}(\tau)$, ${\cal P}(Z)$and  ${\cal P}(\tau^2)$?

\subsection{Transformation of variables}
\label{sec:chvar}
We wish to transform the probability distribution of 
${\cal P}(x)$ to the probability distribution of ${\cal G}(y)$ with $y$ a function of $x$.  Recall that probability is a conserved quantity (we can't create or destroy probabilities...) so 
\begin{equation}
{\cal P}(x)dx={\cal G}(y)dy
\end{equation}
thus 
\begin{equation}
{\cal P}(x)={\cal G}(y(x))\left|\frac{dy}{dx}\right|
\end{equation}

Following the example above if $x$ is $\tau$ and $y$ is $\exp(\tau)$ then ${\cal P}$ is related to ${\cal G}$ by a factor $2 \tau$, and if $y$ is $\tau^2$ by a factor 2. In other words using different priors\index{prior} leads to different posteriors\index{posterior}. This is the main limitation of the Bayesian approach.

\subsection{Marginalization}
\label{sec:marg1}
So far we have considered  probability distributions of a random variable $x$, but one could analogously define {\bf multivariate distributions}\index{multi variate distribution} , the joint probability distribution of two or more variables e.g., ${\cal P}(x,y)$. A typical example is the description of the initial distribution of the density perturbations in the Universe. Motivated by inflation and by the central limit theorem, the initial distribution of density perturbation is usually  described by a multi-variate Gaussian: at every point in space given by its spatial coordinates ($x$, $y$, $z$), ${\cal P}$ is taken to be a random  Gaussian distribution. 
Another example is when one simultaneously constrains the parameters of a model, say for example $\vec{\theta}=\{\Omega_m$, $H_0\}$ (here $H_0$ denotes the Hubble constant).
If you have ${\cal P}(\Omega_m, H_0)$ and want to know the probability distribution of $\Omega_m$ regardless of the values of $H_0$ then:
\begin{equation}
{\cal P}(\Omega_m)=\int dH_0 {\cal P}(\Omega_m, H_0)
\end{equation}

\subsection{Back to statistical inference and Cosmology}
\label{sec:statiinf}
Let us go back to the issue of statistical inference and follow the example from \cite{Wall}.  If you have an urn with $N$ red balls and $M$ blue balls and you draw one ball at the time  then probability theory can tell you what are your chances of picking  a red ball given that you have already drawn $n$ red and $m$ blue: ${\cal P}(D|H)$. However this is not what you want to do: you  want to make a few drawn from the urn and use probability theory to tell you what is the red  vs blue distribution inside the urn: ${\cal P}(H|D)$. In the Frequentist approach  all you can compute is ${\cal P}(D|H)$.

In the case of cosmology it  gets even more complicated. 

We consider that the Universe we live in is a random realization of all the possible 
Universes that could have been a realization of the true underlying model (which is known 
only to Mother Nature). All the possible realizations of this true underlying Universe make 
up the {\it ensamble}. In statistical inference one may sometime want to try to estimate how 
different our particular realization of the Universe could be from the true underlying one. 
Going back to the example of the urn with red and blue balls, it would as if we were to be drawing from one particular urn, but the urn is part of a large batch. On average, the batch distribution has 50\% red and 50\% blue, but each urn has only an odd number of balls so any particular urn cannot reflect exactly the 50-50 spilt.

A  crucial assumption of standard cosmology is that the part of the Universe that we can 
observe is a fair sample of the whole.
But the peculiarity in cosmology is that we have just one Universe, 
which is just one realization from the ensemble (quite fictitious one: it is the ensemble of all 
possible Universes). The fair sample hypothesis states that samples from well separated parts 
of the Universe are independent realizations of the same physical process, and that, in the 
observable part of the Universe, there are enough independent samples to be representative 
of the statistical ensemble.

In addition, experiments in cosmology are not like lab experiments: in many cases observations can't easily be repeated (think about the observation of a particular Supernova explosion or of a Gamma ray burst) and  we can't try to perturb the Universe to see how it reacts... 
After these considerations,  it may be clearer why cosmologists tend to use the Bayesian\index{Bayesian} approach.

\section{Chisquare \& goodness of fit}
\label{sec:3}
Say that you have a set of observations and have a model, described by a set of parameters $\vec{\theta}$, and want to fit the model to the data. The model may be physically motivated or a convenient function.  One then should define a merit function, quantifying the agreement between the model and the data, by maximizing the agreement one obtains the best fit parameters.  
Any useful fitting procedure should provide: 1) best fit parameters  2) estimate of error on the parameters 3) possibly a measure of the goodness of fit. One should bear in mind that if the model is a poor fit to the data then the recovered best fit parameters are meaningless. 

\begin{figure}[t]
\hspace*{-1.cm}
\includegraphics[scale=.37]{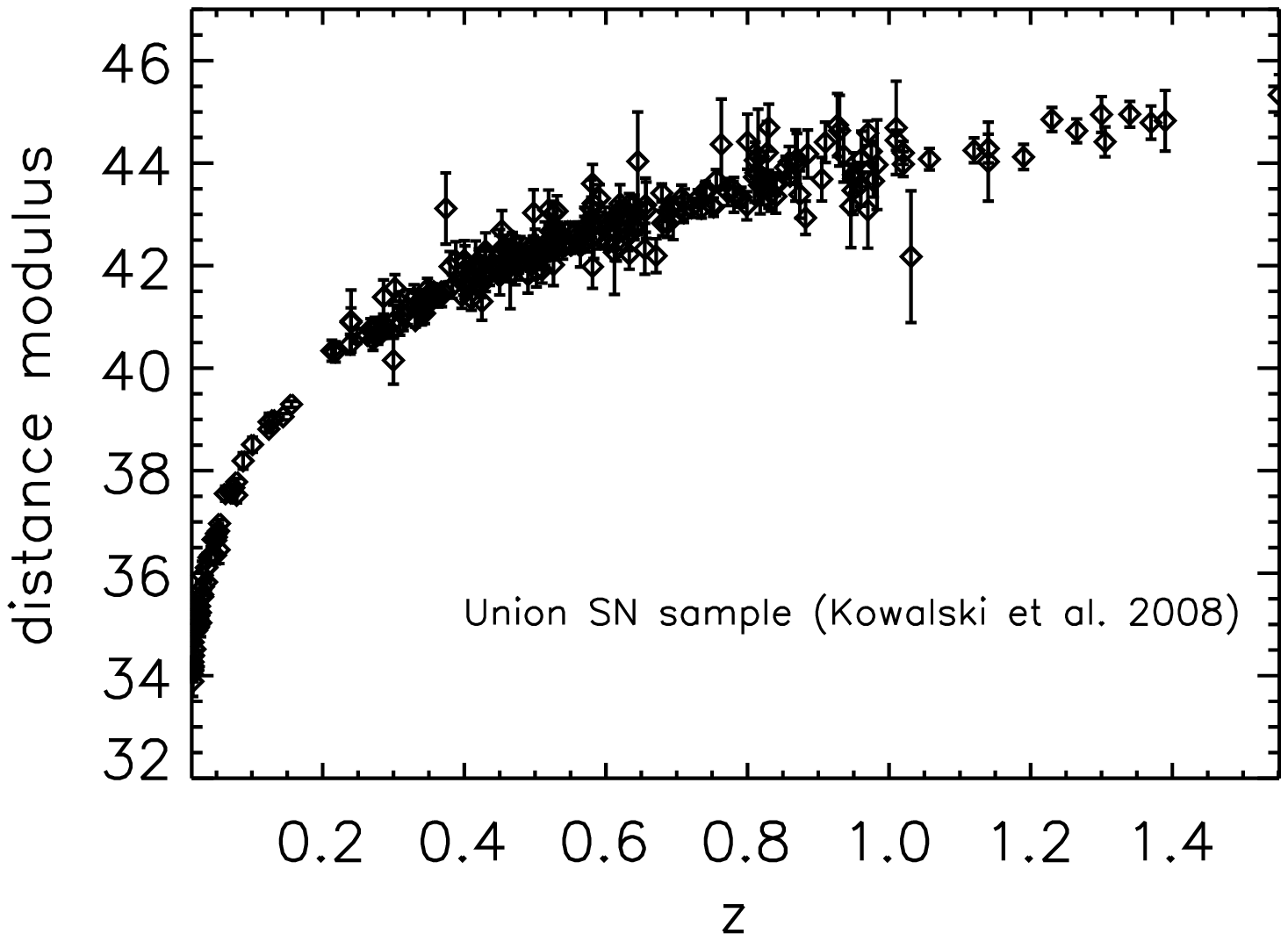}
\includegraphics[scale=.37]{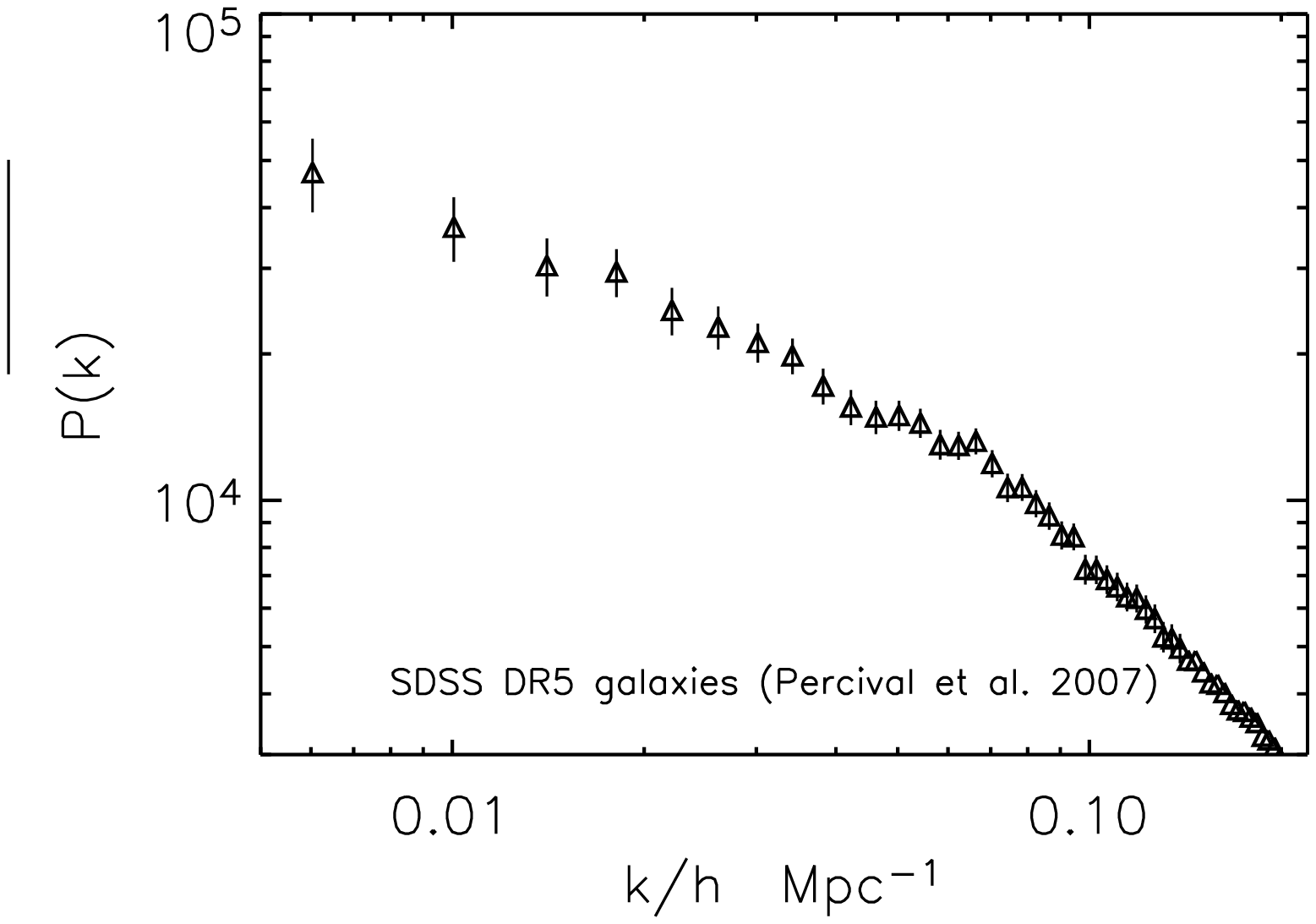}
\caption{Left: distance modulus vs redshift for Supernovae type 1A from the UNion sample \cite{Unionsn}. Right: bandpower  $P(k)$ for DR5 SDSS galaxies, from \cite{percivalDR5}. In both cases one may fit a theory (and the theory parameters) to the data with the Chisquare method. Note that in both cases errors are correlated. In the right panel the errors are also strictly speaking not Gaussianly distributed.}
\label{fig:unionsn}       
\end{figure}

Following Numerical recipes (\cite{numrec}, Chapter 15) we introduce the concept of model fitting (parameter fitting) using least-squares.
Let us assume we have a set of data points $D_i$, for example these could be the band-power galaxy power spectrum at a set of $k$ values, and a model for these data , $y(x,\vec{\theta})$ which depends on  set of parameters $\vec{\theta}$ (e.g. the LCDM power spectrum, which depends on $n_s$--primordial power spectrum spectral slope-, $\sigma_8$ --present-day amplitude of  rms mass fluctuations on scale of 8 Mpc/h--, $\Omega_m h$ etc.). Or it could be for example the supernovae type 1a distance modulus as a function of redshift see e.g. Fig.(\ref{fig:unionsn}) \cite{Unionsn,percivalDR5}.

The least squares, in its simplest incarnation  is:
\begin{equation}
\label{ceq:hisq1}
\chi^2=\sum_i w_i [D_i-y(x_i|\vec{\theta})]^2
\end{equation}
where $w_i$ are suitably defined weights.
It s possible to show that the minimum variance weight is $w_i=1/\sigma_i^2$ where $\sigma_i$ denote the error on data point $i$. In this case then the least squares is called Chisquare\index{Chisquare}.
If the data are correlated the chisquare becomes:
\begin{equation}
\label{eq:chisq2}
\chi^2=\sum_{ij} (D_i-y(x_i|\theta)Q_{ij}(D_j-y(x_j|\theta)
\end{equation}
where $Q$ denotes the inverse of the so-called covariance matrix describing the covariance between the data.
The best fit parameters are those that minimize the $\chi^2$. See an example in Fig.\ref{fig:chisqexample}.

For a wide range of cases the probability distribution for different values of $\chi^2$ around the minimum  of Eq.\ref{eq:chisq2} is the $\chi^2$ distribution  for $\nu=n-m$ degrees of freedom where $n$ is the number of independent  data  points and $m$ the number of parameters. The probability that the observed $\chi^2$ exceeds by chance a value $\widehat{\chi}$ for the correct model  is $Q(\nu,\widehat{\chi})=1-\Gamma(\nu/2,\widehat{\chi}/2)$ where $\Gamma$ denotes the incomplete Gamma function. See  the Numerical Recipes bible \cite{numrec}. Conversely, the probability the the observed $\chi^2$, even for the correct model, is less than  $\widehat{\chi}$ is $1-Q$.  While this statement is strictly true if measurement errors are Gaussian and the model is a linear function of the parameters, in practice it applies to a much wider rangeof cases.

The quantity $Q$ evaluated the the minimum chisquare  (i.e. at the best fit values for the parameters) give a measure of the goodness of fit.  If $Q$ gives a very small probability then there are three possible explanations:\\
1) the model is wrong and can be rejected. (Strictly speaking: the data are unlikely to have happened  if the Universe was really described by the model considered) \\
2) the errors are underestimated\\
3) the measurement errors are non Gaussianly distributed. \\

Note that in the example of the power spectrum we know a priori that the errors are not Gaussianly distributed. In fact,  even if the initial conditions were Gaussian and if the underlying matter perturbations  were  evolving  still in the linear regime (i.e.$\delta \rho/\rho \ll 1$) and galaxies were nearly unbiased tracers of the dark matter, then the density fluctuation itself would obey Gaussian statistics and so would its Fourier transform, but {\bf not} its power spectrum, which is a square quantity.
In reality  we  now that by $z=0$ perturbations grow non-linearly and that galaxies may not be nearly unbiased tracers of the underlying density field. Nevertheless, the Central Limit theorem comes to our rescue, if in each band-power there is a sufficiently large number of modes.

If $Q$ is too large (too good to be true)  it is also cause for concern:\\
1) errors have been overestimated\\
2) data are correlated or non-independent\\
3) the distribution is non Gaussian\\
Beware: this last case  is very rare.

A useful ``chi-by-eye" rule is: the minimum $\chi^2$ should be roughly equal to $\nu$ (number of data - number of parameters). This is increasingly true for large $\nu$. From this, it is easy to understand the use of the so-called  ``reduced chisquare" that is the $\chi^2_{min}/m$: if $m \gg n$ (i.e., number of data much larger than the number of parameters to fit, which should be true in the majority of the cases)  then $m \sim \nu$ and the rule of thumb is that  reduced chisquare should be unity. 

Note that the chisquare method, and the $Q$ statistic, give the probability for the data given a model ${\cal P}(D|\theta)$, not ${\cal P}(\theta|D)$. One can make this identification via the prior. 

\section{Confidence regions}
\label{sec:4}
Once the best fit parameters are obtained, how can one represent the confidence limit or confidence region around the best fit parameters? A reasonable choice is to find a region in the $m$-dimensional parameter space (remember that $m$ is the number of parameters) that contain a given percentage of the probability distribution. In most cases one wants a compact region around the best fit vales. 
A natural choice is then given by regions of constant $\chi^2$\index{Chisquare} boundaries.
Note that there may be cases (when the $\chi^2$ has more than one minimum)  in which one may need to report a non-connected confidence region. For multi-variate Gaussian distributions  however these are  ellipsoidal regions. Note that the fact that the data have Gaussian errors does not imply that  the parameters will have a Gaussian probability distribution...
\begin{figure}[t]
\hspace*{-1.cm}
\includegraphics[scale=.37]{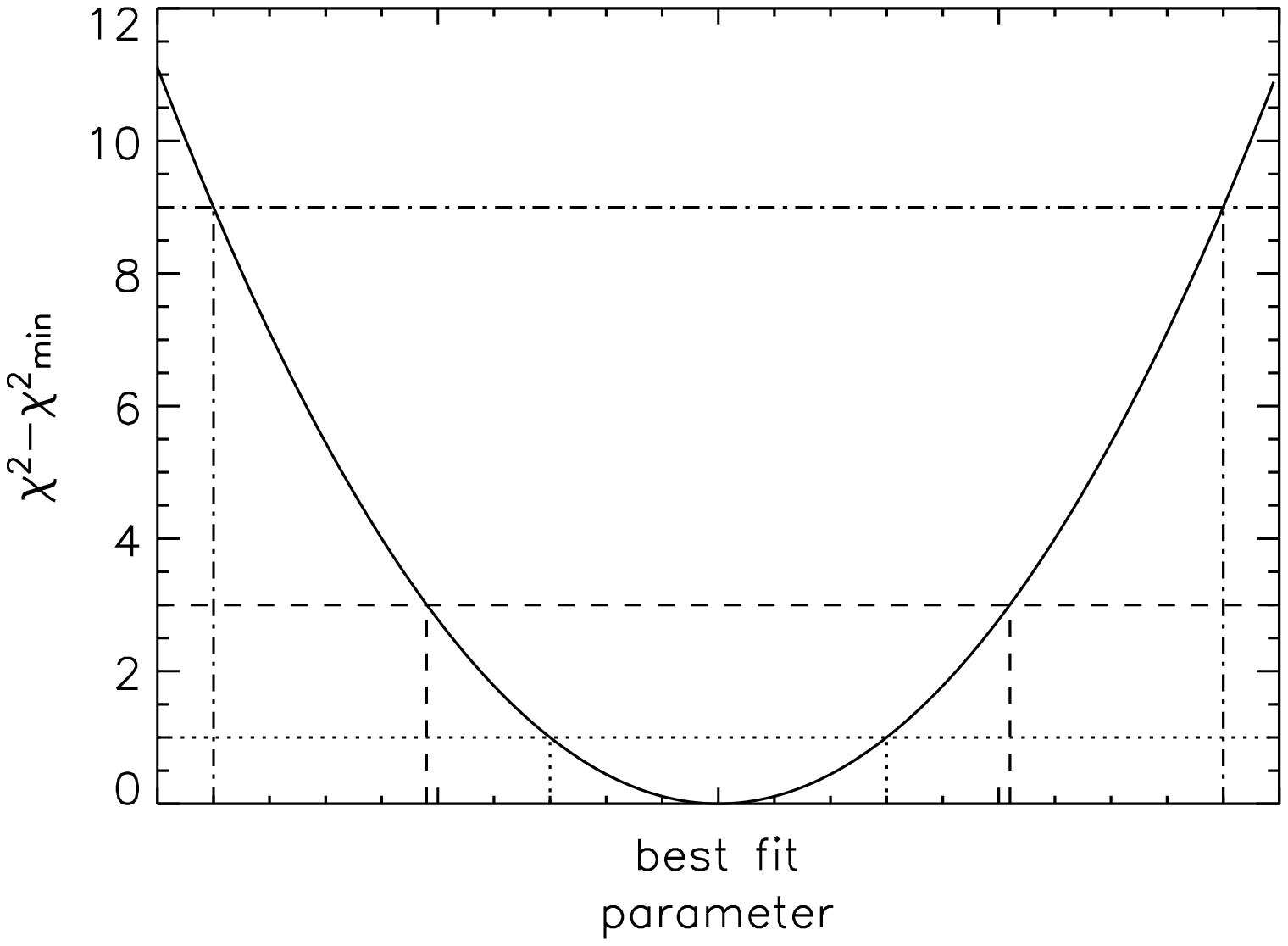}
\includegraphics[scale=.33]{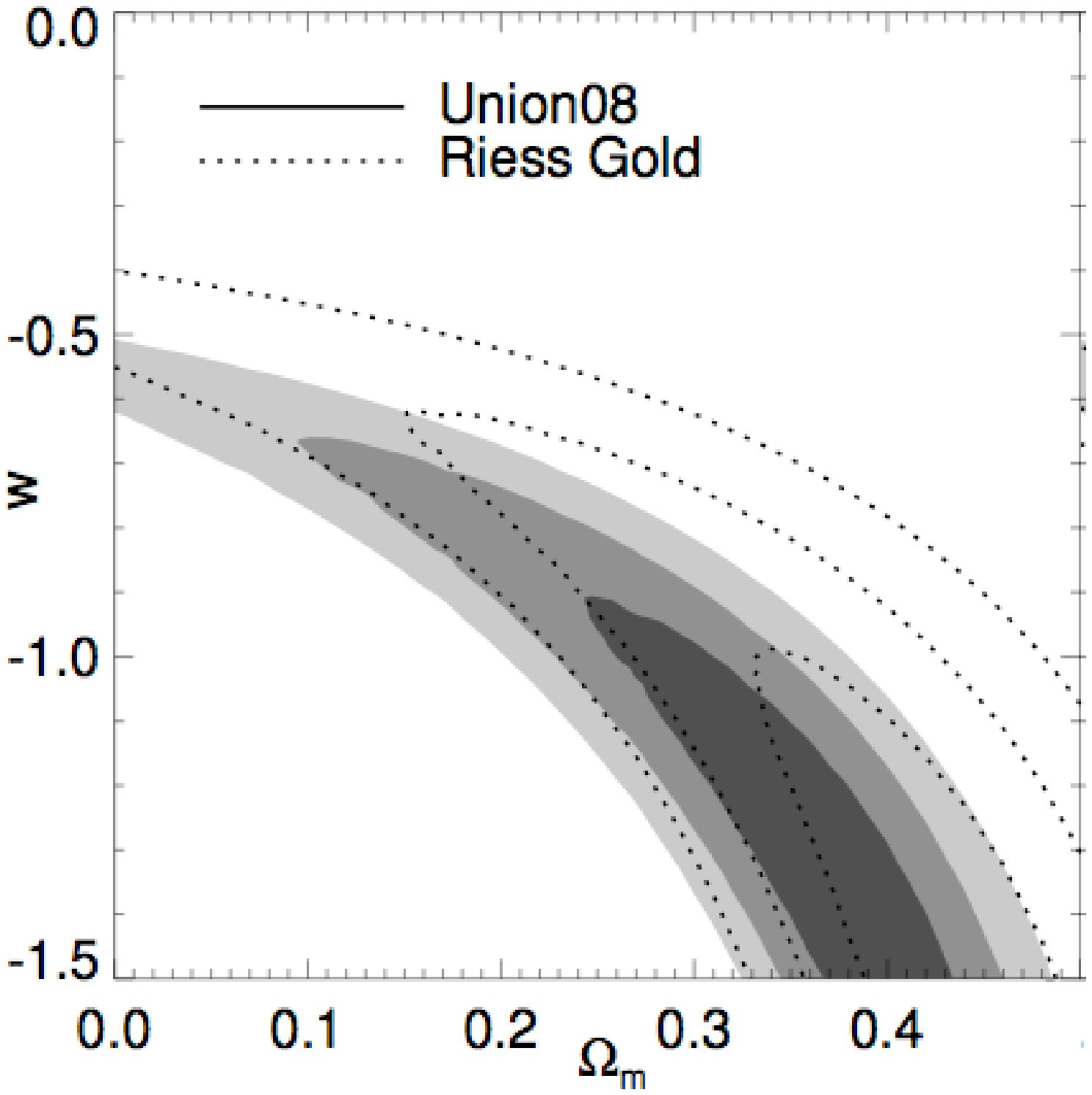}
\caption{Left: example of a one-dimensional chisquare for a Gaussian distribution as a function of a parameter and corresponding  68.3\%, 95.4\% and 99.5\% confidence levels. Right a two dimensional example for the Union supernovae data. Figure from Kowlaski et al. (2009)\cite{Unionsn} reproduced by permission of the AAS. Note that in a practical application even if the data have gaussian errors the errors on the parameter may not be well described by multi-variate Gaussians (thus the confidence regions are not ellipses).}
\label{fig:chisqexample}       
\end{figure}

Thus, if the values of the parameters are perturbed from the best fit, the $\chi^2$ will increase.  One can use the properties of the $\chi^2$ distribution to  define confidence intervals in relation to $\chi^2$variations or $\Delta \chi^2$.
Table 1  reports the $\Delta \chi^2$ for 68.3\%, 95.4\% and 99.5\% confidence levels as function of number of parameters for the joint confidence level. In the case of Gaussian distributions these correspond to the conventional 1, 2 and 3 $\sigma$. Se an example of this in Fig.\ref{fig:chisqexample}

Beyond these values here is the general prescription to compute  constant-$\chi^2$ boundaries confidence levels.
After having found the best fit parameters by minimizing the $\chi^2$ and if $Q$ for the best fit parameters  is acceptable then: \\
\begin{itemize}
\item{1} Let $m$ be the number of parameters, $n$ the number of data and $p$ be the confidence limit desired.
\item{2} Solve the following equation for $\Delta \chi^2$:
\begin{equation}
Q(n-m,min(\chi^2)+\Delta \chi^2)=p 
\end{equation}
\item{3} Find the parameter region where $\chi^2\le min(\chi^2)+\Delta \chi^2$. This defines the confidence region.
\end{itemize}
\begin{table}
\caption{$\Delta \chi^2$ for  the conventionals $1,2$,and $3-\sigma$ as a function of the number of parameters for the joint confidence levels.}
\label{table:chisq}
\begin{tabular}{|c|c|c|c|}
\hline
p&1&2&3\\
\hline
68.3\%&1.00&2.30&3.53\\
95.4\%&2.71&4.61&6.25\\
99.73\%&9.00&11.8&14.2\\
\hline
\end{tabular}
\end{table}

If the actual error-distribution is non Gaussian but it is known then it is still possible to use the $\chi^2$ approach, but instead of using the chisquare distribution and table \ref{table:chisq}, the distribution need to be calibrated on multiple simulated realization of the data as illustrated below in the section dedicated to Monte Carlo  methods.

\section{Likelihood}
\label{sec:5}
So far we have dealt with the frequentist quantity ${\cal P}(D|H)$. If we set ${\cal P}(D)=1$ and ignore the prior then we can identify the likelihood with ${\cal P}(H|D)$ and thus by maximizing the likelihood\index{likelihood} we can find the most likely model (or model's parameters) given the data. However having ignored  ${\cal P}(D)$ and  the prior this approach cannot give in general a goodness of fit and thus cannot give an absolute probability for a given model. However it can give relative probabilities.
 If the data are Gaussianly distributed the likelihood is given by a multi-variate Gaussian:
 \begin{equation}
 {\cal L}=\frac{1}{(2\pi)^{n/2} |det C|^{1/2}} \exp\left[-\frac{1}{2} \sum _{ij}(D-y)_iC^{-1}_{ij}(D-y)_j\right]
 \end{equation}
where $C_{ij}=\langle (D_i-y_i)(D_j-y_j)\rangle$ is the covariance matrix.

It should be clear from this that the relation between $\chi^2$\index{Chisquare} and likelihood is that, for Gaussian distributions, 
${\cal L}\propto \exp[-1/2 \chi^2]$ and minimizing the $\chi^2$ is equivalent at minimizing the likelihood. In this case likelihood analysis and $\chi^2$ coincide and by the end of this section,  it will this be no surprise that the Gamma function appearing in the $\chi^2$ distribution is closely related to the Gaussian integrals.

The subtle step is that now, in Bayesian\index{Bayesian} statistics, confidence regions are regions $R$ in {\it model space} such that
$\int_R {\cal P}(\theta|D) d\theta =p$ where $p$ is the confidence level we request (e.g., 68.3\%, 95.4\% etc.). Note that by integrating the posterior\index{posterior} over the model parameters, the confidence region depends on the prior information: as seen in \S \ref{sec:chvar} different priors give different posteriors and thus different regions $R$.

It is still possible to report results independently of the prior by using the {\it Likelihood ratio}\index{likelihood ratio}. The likelihood at a particular point in parameter space is compared with that at  the best fit value, ${\cal L}_{max}$ where likelihood id maximized. 
Thus a model is acceptable if  the likelihood ratio
\begin{equation}
\Lambda=-2 \ln \left[\frac{ {\cal L}(\theta)}{{\cal L}_{max}}\right]\,
\end{equation}
is above a given threshold.
The connection to the $\chi^2$ for Gaussian distribution should be clear. In general, the threshold can be calibrated by calculating the entire distribution of the likelihood ratio in the case that a particular model is the true model. Frequently this is chosen to be the  best ft model.

There is a subtlety to point out here. In cosmology the data may be Gaussainly distributed and still the $\chi^2$ and likelihood ratio analysis may give different results. This happens because in identifying likelihood and chisquare we have neglected the term $[(2\pi)^{n/2} |det C|^{1/2}]^{-1}$. If the covariance does not depend on the model or model parameters, this is just a normalization factor which drops out in the likelihood ratio. However in cosmology often the covariance depends on the model: this happens for example when errors are dominated by cosmic variance, like in the case of the CMB  temperature fluctuations  on the largest scales, or on the galaxies power spectrum on the largest scales. 
In this case the cosmology dependence of the covariance cannot be neglected, but  one can always define a pseudo-chisquare as $-2 \ln {\cal L}$ and work with this quantity. 

Let us stress again that the likelihood is linked to the posterior\index{posterior} through the prior: the identification of the likelihood with the posterior is prior dependent (as we will see in an example below).
In the absence of any data it is common to assume a flat (uniform) prior: i.e. all value of the parameter in question are equally likely, but other choices are possible and sometimes more motivated.  For example, if a parameter is positive-definite, it may be interesting to use a logaritmic prior (uniform in the $\log$).

Priors may be assigned theoretically or  from prior information gathered from previous experiments.  If the priors are set by theoretical considerations, it is always good practice to check how much the results depend on the choice of the prior. If the dependence is significant, it means that the data do not have much statistical power to constrain that (those) parameter(s). Information theory helps us quantify the amount of ``information gain" : the information in the posterior relative to the prior:
\begin{equation}
{\cal I} =\int {\cal P}(\theta|D) \log \left[\frac{{\cal P}(\theta|D)}{{\cal P}(\theta)}\right]d\theta
\end{equation}

\subsection{Marginalization: examples}
\label{sec:marg}
 Some of the model parameters  may be uninteresting.  For example, in many analyses one wants to include nuisance parameters (calibration factors, biases, etc.) but then report the confidence level on the real cosmological parameters regardless of the value of the nuisance ones.  In other cases the model may have say, 10 or more real cosmological parameters but we may be interested in the  allowed range of only one or two of them, regardless of the values of all  the  other. Typical examples are e.g., constraints on the curvature parameter $\Omega_k$ (which we may want to know regardless of the values of e.g.,  $\Omega_m$ or $\Omega_{\Lambda}$) or, say, the allowed range for the neutrino mass regardless of the power spectrum spectral index or the value of the Hubble constant. As explained in \S\ref{sec:marg1} one can marginalize over the uninteresting parameters.
 
 It should be kept in mind that marginalization is a Bayesian\index{Bayesian} concept: the results may depend on the prior chosen.
 
 In some cases, the marginalization can be carried out analytically. An example is reported below: this applies to the case of e.g., calibration uncertainty, point sources amplitude, overall scale independent galaxy bias, magnitude intrinsic brightness or beam errors for CMB studies. In this case it is useful to know the following results for Gaussian likelihoods\index{likelihood}:
 
\begin{eqnarray}
{\cal P}(\theta_1..\theta_{m-1}|D)\!\!\!\!&=&\!\!\!\!\!\!\int \frac{dA}{(2\pi)^{\frac{m}{2}} ||C||^{\frac{1}{2}}} e^{\left[-\frac{1}{2}(C_i-(\hat{C_i}+AP_i))\Sigma_{ij}^{-1}(C_j-(\hat{C_j}+AP_j))\right]}  \\ \nonumber
&\times&\frac{1}{\sqrt{2\pi \sigma^2}}\exp\left[-\frac{1}{2}\frac{(A-\hat{A})^2}{\sigma^2}\right]         
\end{eqnarray}
repeated indices are summed over and $||C||$ denotes the determinant. Here, $A$ is the amplitude of, say, a point source contribution $P$ to the $C_{\ell}$ angular power spectrum, $A$ is the  $m^{th}$ parameter which we want to marginalize over  with a  Gaussian prior with variance $\sigma^2$ around $\hat{A}$.
The trick is to recognize that this    integral can be written as:
\begin{equation}
\!\!\!\!{\cal P}(\theta_1..\theta_{m-1}|D)=C_0 \exp\left[-\frac{1}{2}C_1- 2 C_2 A+C_3 A^2\right]dA
\end{equation}
(where $C_{0...3}$ denote constants and it is left as an exercise to  write them down  explicitly)
and that  this kind of integral  is evaluated by using the substitution $A \longrightarrow A-C_2/C_3$ giving something $\propto \exp[-1/2(C_1 - C_2^2/C_3)]$.

\begin{figure}[t]
\includegraphics[scale=.38]{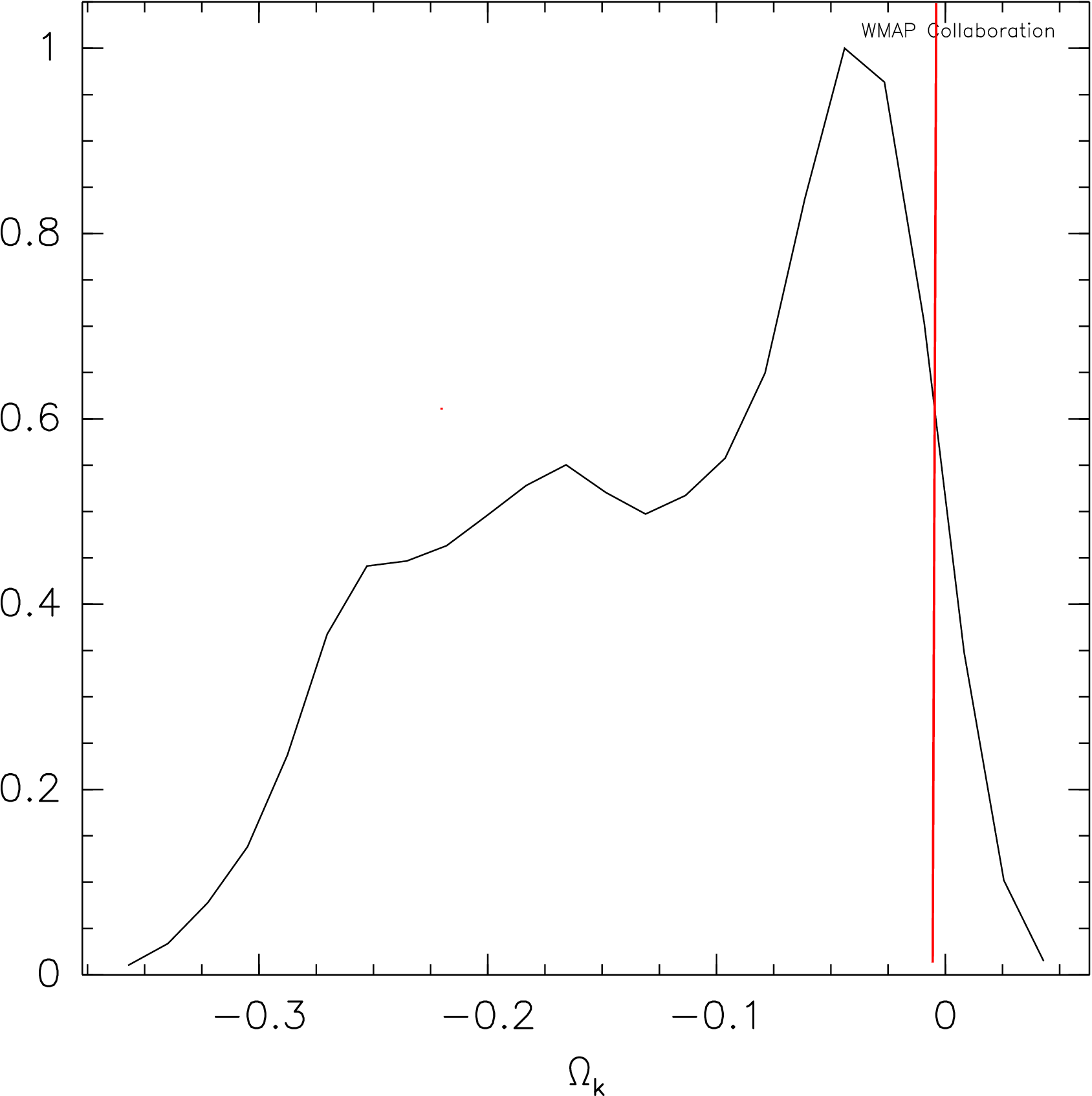}\\
\includegraphics[scale=.75]{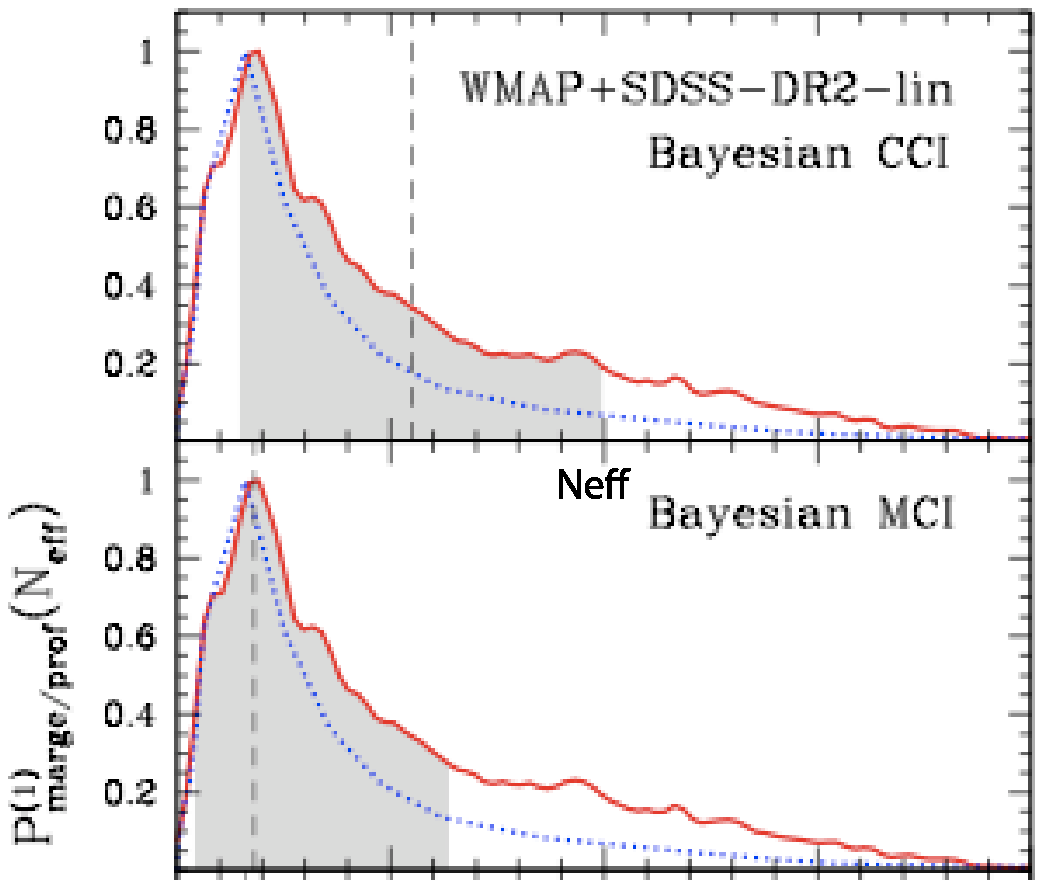}
\caption{Marginalization effects. Top panel: We consider the posterior distribution for the cosmological parameters of a dark energy + cold dark matter model where curvature is a free parameter and so is a (constant) equation of state parameter for dark energy.  The data are the WMAP 5 year data. The red line shows the N-dimensional maximum posterior value and the black line is the marginalized posterior over all other cosmological parameters. Figure courtesy of LAMBDA \cite{Lambda}. Bottom panel: figure from \cite{haman07}. Illustration of Central Credible Interval (CCI) and Minimum Credible Interval (MCI), for the case of a LCDM model with free number of effective neutrino species (ignore blue dotted line for this example, red line is the marginalized posterior).  }
\label{fig:omegak}       
\end{figure}

In cases where the likelihood surface (describing the value of the likelihood as a function of the parameters) is not a multi-variate Gaussian, the location of the maximum likelihood before marginalization may not coincide with the location after marginalization. An example is shown in Fig.\ref{fig:omegak}. The figure show the probability distribution for $\Omega_k$ form WMAP5  data for a model where curvature is free and the equation of state parameter for dark energy $w$ is constant in time but not fixed at $-1$.  The red line shows the N-dimensional maximum posterior value and the black line is the marginalized posterior over all other cosmological parameters.

It should also be added that, even in the case where we have a single-peaked posterior probability distribution  there are  two common estimators of the ``best" parameters: the peak value (i.e. the most probable value) or the mean, $\hat{\theta}=\int d\theta \theta {\cal P}(\theta|D)$.
If the posterior is non-Guassian these two estimates need not to coincide. In the same spirit, slightly different definitions of confidence intervals   need not to coincide for non-Gaussian likelihoods, as illustrated in the right panel of Fig. \ref{fig:omegak}: for example one can define the confidence interval $[\theta_{low},\theta_{high}]$, such that  equal fractions of the posterior volume lie in ($-\infty, \theta_{low}$) and ($\theta_{high}, \infty$). This is called central credible interval and is connected to the median.
Another possibility (minimum credible interval) is to consider the region so that the posterior at any point inside it is larger that at any point outside and so that  the integral of the posterior in this region is the required fraction of the total. Thus remember, it is always good practice to declare what confidence interval one is using!
This subject is explored in more details in e.g.,  \cite{haman07}.

\section{Why Gaussian Likelihoods?}
Throughout this lectures we always refer to Gaussian likelihoods. It is worth mentioning that if the data errors are Gaussianly distributed then the likelihood function for the data will be a multi-variate Gaussian.  If the data are not Gaussianly distributed (but still are drawn from a distribution with finite variance!) we can resort to the central limit theorem:  we can bin the data  so that in each bin there is a superposition of many independent measurement. The central limit theorem will tell us that the resulting distribution (i.e. the error distribution for each bin) will be better approximated by a multi-variate Gaussian. 
However, as mentioned before,  even if the data are Gaussianly distributed this does not ensure that the likelihood surface for the parameters will be a multi-variate Gaussian: for this to be  always true   the model needs to depend linearly on the parameters.   
Even without resorting to the central limit theorem, the Gaussian approximation is in many cases recovered even when starting from highly non-gaussian distribution. A neat example is provided by  Cash \cite{Cash} \index{Cash statistics} which we follow here.

Let's say you want to constrain cosmology by studying clusters number counts as a function of redshift. 
The observation of a discrete number $N$ of clusters is  a Poisson process, the probability of which is given by the product 
\begin{equation}
{\cal P}=\Pi_{i=1}^{N}[e_i^{n_i}\exp(-e_i)/n_i!]
\end{equation}
 where $n_i$ is the number of clusters observed in the $i-th$
 experimental bin and $e_i$ is the expected number in that bin in a
 given model: $e_i=I(x)\delta x_i$ with $i$ being the proportional to
 the probability distribution.  Here $\delta x_i$ can represent an interval in clusters mass and/or redshift.
 Note: this is a product of Poisson distributions, thus one is assuming that these are independent processes. Clusters may be clustered, so when can this be used?
 
 For unbinned data (or for small bins so
 that bins have only 0 and 1 counts) we define the quantity:
\begin{equation}
C\equiv -2 \ln {\cal P}=2(E-\sum_{i=1}^N\ln I_i)
\end{equation}
where $E$ is the total expected number of clusters in a given  model.
The quantity $\Delta C$ between two models with different parameters has a $\chi^2$ distribution! (so all  that was said in the $\chi^2$ section applies, even though we started from a highly non-Gaussian distribution.)  

\section{The effect of priors: examples}
\label{sec:6}
\begin{figure}[t]
\includegraphics[scale=.7,viewport=30 10 800 300,clip]{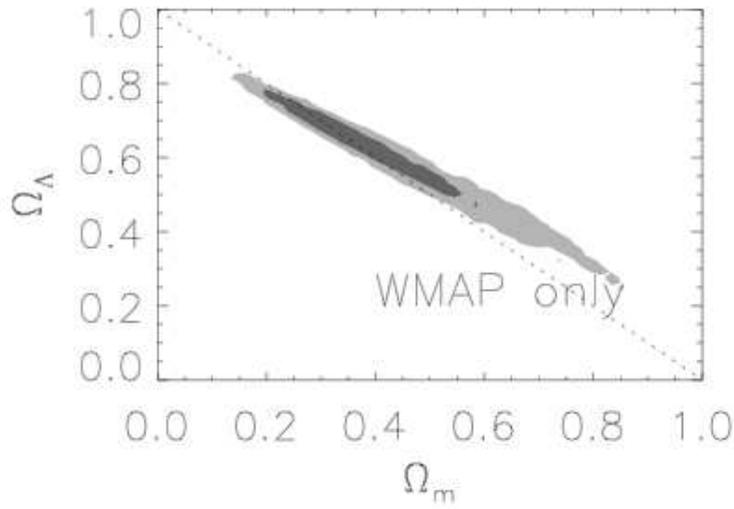}
\includegraphics[scale=.5,viewport=-30 -120 800 300,clip]{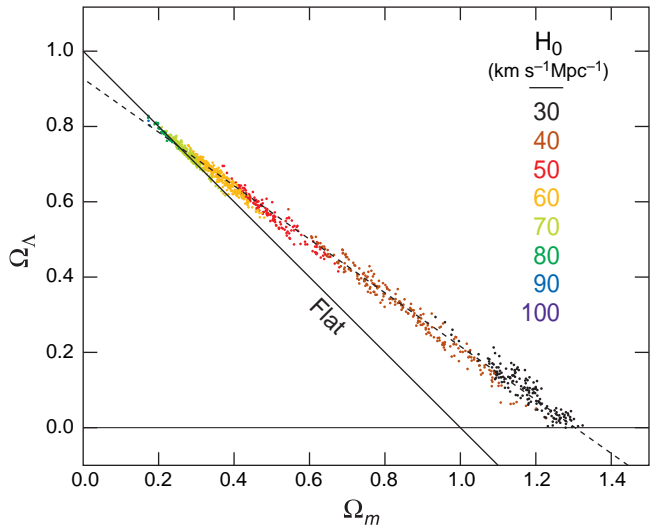}
\caption{Top: WMAP 1st year data constraints in the $\Omega_m$, $\Omega_{\Lambda}$ plane, from  Spergel et al 2003, ApJS, 148:175-194 \cite{wmap1params}. Bottom: models consistent with the WMAP 3 yr data, from Spergel et al. (2007) ApJS, 170, 377\cite{wmap3params}. In both cases the model is a non-flat LCDM model. Figures reproduced by permission of the AAS.}
\label{fig:priors}       
\end{figure}

Let us consider the two  figures in Fig. \ref{fig:priors}. On the left: WMAP 1st year data constraints in the $\Omega_m$, $\Omega_{\Lambda}$ plane. On the right: models consistent with the WMAP 3 yr data.  In both cases the model is a non-flat LCDM model. So why the addition of more data (the  two extra  years of WMAP observations) gives worst constraints?
The key is that what is reported in the plots is a representation of the posterior probability distribution. 
In the left panel a flat prior on $\Theta_A$ (angular size distance to the last scattering surface,  giving by the position of the first peak) was assumed. In the figure on the right a flat prior on the Hubble constant $H_0$ was assumed. Remember: always declare the priors assumed!
 
\section{Combining different data sets: examples}
\label{sec:combine}
\begin{figure}[t]
\includegraphics[scale=.57]{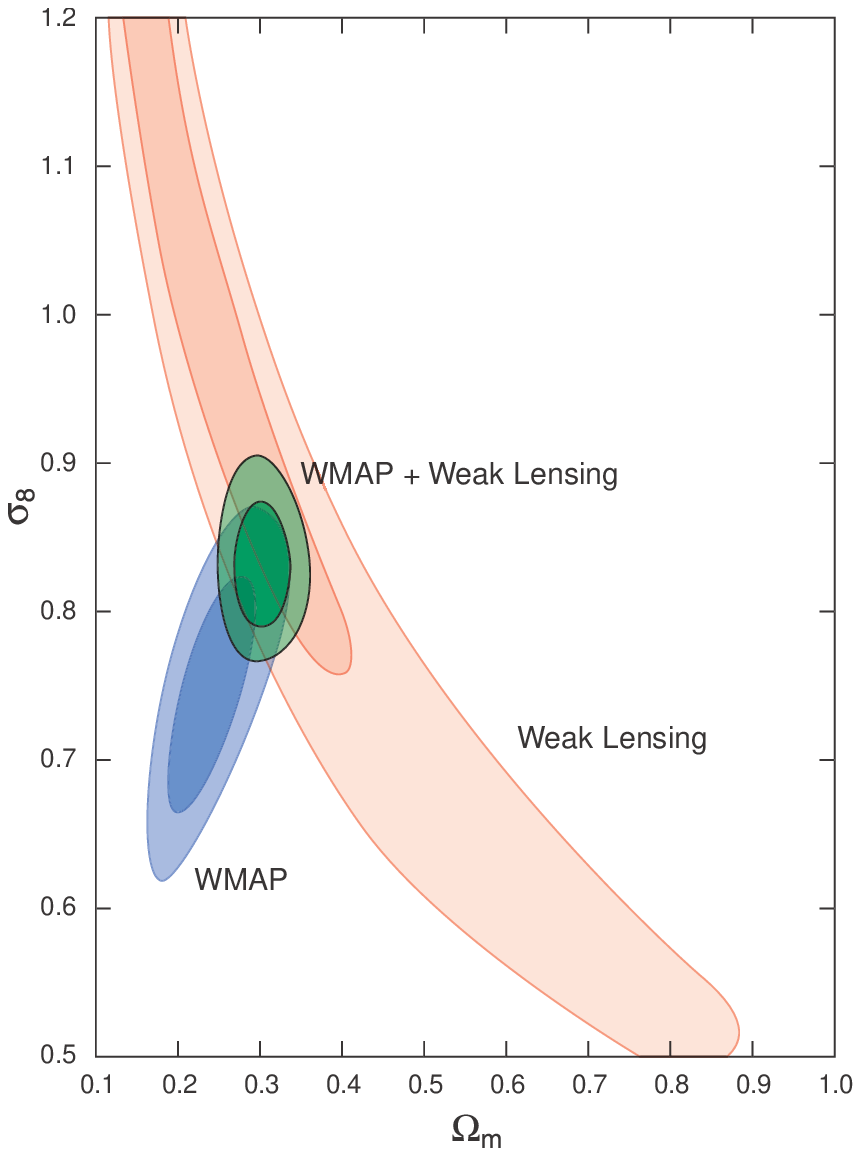}
\includegraphics[scale=.38]{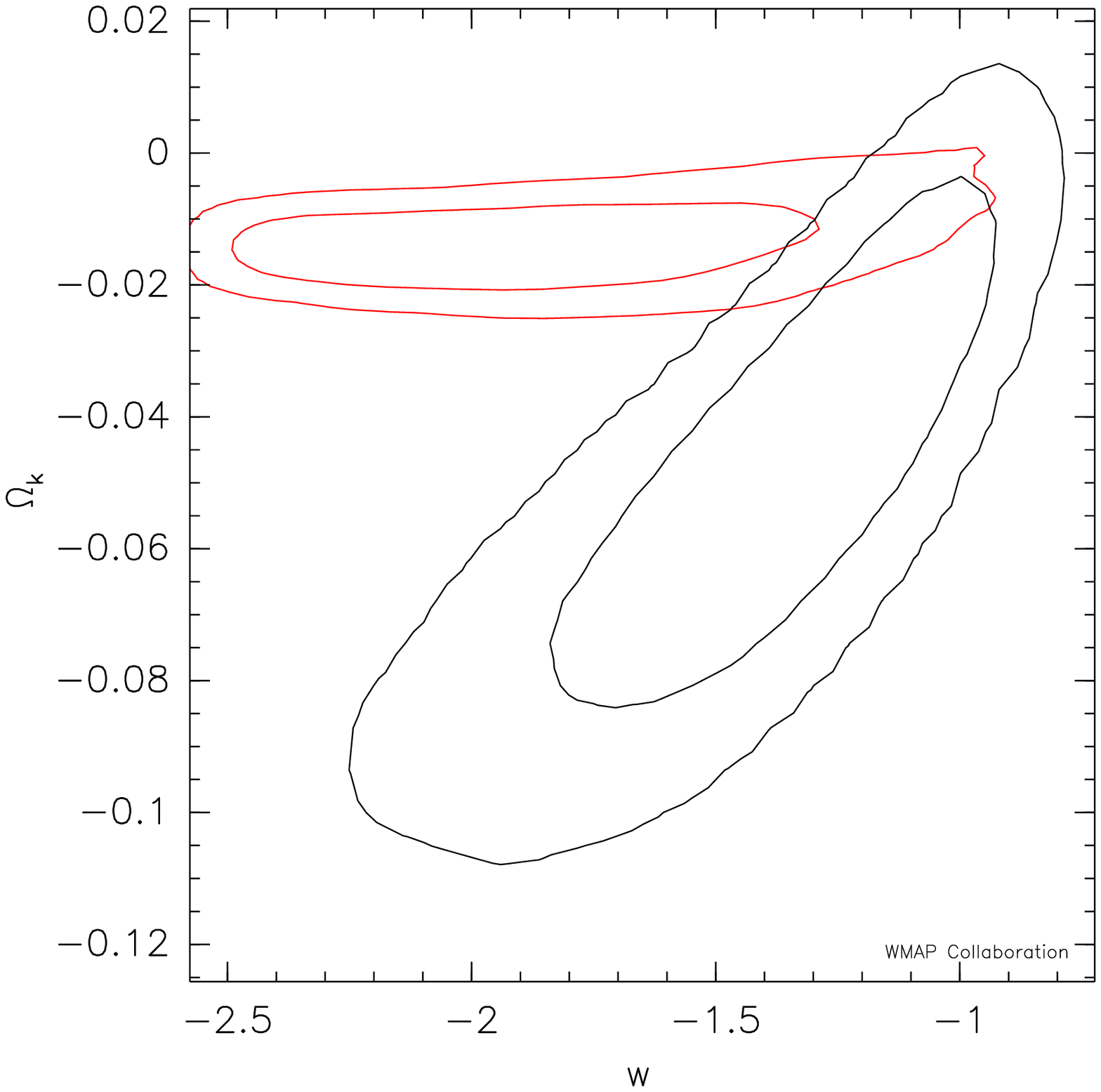}
\caption{Left: constraints in the $\Omega_m$ $\sigma_8$ plane for a flat LCDM model for WMAP 3yr data (blue),  weak lensing constraints (orange) and combined constraints.Figure from Spergel et al. 2003 \cite{wmap3params}, reproduced by permission of the AAS. Right:Constraints in the $\Omega_k,w$ plane for  non-flat dark energy models with constant $w$ for WMAP5+ supernovae data (in black) and WMAP5+BAO (in red). Figure courtesy of LAMBDA \cite{LAMBDA}.}
\label{fig:inconsistent}       
\end{figure}

It has become common to ``combine data sets" and explore the constraints from the "data set combination". This means in practice that the likelihoods can be multiplied if the data sets are independent (if not the one should account for the appropriate covariance).
it is important to note that: {\it If the data-sets are inconsistent, the resulting constraints from the combined data set are nonsense}.
An example is shown in Fig.~\ref{fig:inconsistent}. 

On the left panel  we show a figure from \cite{wmap3params} constraints in the $\Omega_m$, $\sigma_8$ plane for a flat LCDM model for WMAP 3yr data (blue), weak lensing constraints (orange) and combined constraints. 
On the right panel the figure shows the constraints in the $\Omega_k,w$ plane for  non-flat dark energy models with constant $w$ for WMAP5+ supernovae data (in black) and WMAP5+BAO (in red). Even though the WMAP data are in common there is some tension in  the resulting constraints. The two data sets (Supernovae and BAO, WMAP and weak lensing ) are not  fully consistent: as the authors themselves, note, they should not be  combined.
 
\section{Forecasts: Fisher matrix}
\label{sec:7}


Before diving into the details let us re-examine  of error estimates for parameters from the likelihood. Let us assume a flat prior in the parameter so we can identify the posterior with the likelihood.
Close to the peaks we can expand the log likelihood in Taylor series:
\begin{equation}
\ln {\cal L}=\ln{\cal L}(\theta_0)+\frac{1}{2}\sum_{ij}(\theta_i-\theta_{i,0})\left.\frac{\partial^2\ln{\cal L}}{\partial \theta_i \partial\theta_j}\right|_{\theta_0}(\theta_j-\theta_{j0})+...
\end{equation}
by truncating this expansion to the quadratic term (remember that by expanding around the maximum we have the first derivative equal to  zero) we say the the likelihood surface is locally a multi-variate Gaussian.
The Hessian matrix is defined as
\begin{equation}
{\cal H}_{ij}=-\frac{\partial^2\ln{\cal L}}{\partial \theta_i \partial\theta_j}.
\end{equation}
It enclose information on the parameters errors and their covariance. If this matrix is not diagonal it means that the parameters estimates are correlated. Loosely speaking we said ``the parameters are correlated": it means that they have a similar effect on the data and thus the data have hard time in telling them apart. The parameters may or may not be physically related with each other.

More specifically if all parameters are kept fixed except one (parameter $i$,say), the error on that parameter would be given by $1/\sqrt{{\cal H}_{ii}}$. This is called conditional error but is almost never used or interesting.

Having understood this, we can move on  to the Fisher\index{Fisher matrix} information matrix \cite{Fisher35}. The Fisher matrix plays a fundamental role in  forecasting errors  from a given experimental set up and thus is the work-horse of experimental design.
It is defined as:
\begin{equation}
\label{eq:fisher}
F_{ij}=-\left\langle \frac{\partial ^2 \ln {\cal L}}{\partial \theta_i \partial \theta_j}\right\rangle
\end{equation} 
It should be clear that  $F=\langle {\cal H}\rangle$.

Here the average is the ensamble average over observational data (those that would be gathered if the real Universe was given by the model --and model parameters-- around which the derivative is taken).
 Since, as we have seen the likelihood for independent data sets is the product of the likelihoods, it follows that the Fisher matrix for independent data sets is the sum of the individual Fisher matrices. This will become useful later on.
 
In the one-parameter case-say only $i$ component of $\theta$, thinking back at the Taylor expansion around the maximum of the likelihood we have that 
\begin{equation}
\Delta \ln {\cal L}=\frac{1}{2}F_{ii} (\theta_i-\hat{\theta_i})^2
\end{equation} 
when $2\Delta \ln {\cal L}=1$ and by identifying it with the $\Delta \chi^2$ corresponding to $68\%$ confidence level, wee see that $1/\sqrt{F_{ii}}$ yields the $1-\sigma$ displacement for $\theta_i$.  This is the analogous to the conditional error from above.
In the general case:
\begin{equation}
\sigma^2_{ij}\ge (F^{-1})_{ij}.
\end{equation}
Thus  when all parameters are estimated simultaneously from the data  the marginalized error is
\begin{equation}
\sigma_{\theta_i}\ge(F^{-1})_{ii}^{1/2}
\end{equation}

Let's spell it out for clarity:  this is the square root of the element $ii$ of the inverse of the Fisher information matrix\footnote{i.e. you have to perform a matrix inversion first.}.  This assumes that the likelihood is a Gaussian around its maximum (the fact that the data are Gaussianly distributed is no guarantee that the likelihood will be Gaussian, see e.g. Fig\ref{fig:chisqexample}).
The terrific utility of the Fisher\index{Fisher matrix} Information matrix is that, if you can compute it,  it enables you to estimate the parameters errors {\bf  before you do the experiment}. If it can be compute it quickly, it also  enables one to explore different experimental set ups and optimize the experiment. This is  why the Fisher matrix approach is so useful in survey design. Also  complementarity of  different, independent and uncorrelated experiments (i.e. how in combination they can lift degeneracies)  can be quickly explored: the combined Fisher matrix is the sum of the individual matrices.  This is of course extremely useful, however read below for some caveats.

The $\ge$ is the Kramer-Rao inequality: the Fisher\index{Fisher matrix} matrix approach always gives you an optimistic estimate of the errors (reality is only going to be worst). And this is not only because systematic and real world effects are often ignored in the Fisher information matrix calculation, but for a fundamental limitation: only if the likelihood is Gaussian that $\ge$ becomes $=$. 
In some cases, when the Gaussian approximation for the Likelihood does not hold, it is possible to make non-linear transformation of the parameter that make the likelihood Gaussian. Basically, if the data and Gaussianly distributed and the model depends linearly on the parameters then the likelihood would be Gaussian. So the key is to have a good enough understanding of the theoretical model to be able to find such a transformation. See \cite{normalparams} for a clear example.

\begin{figure}[t]
\hspace*{-1.cm}
\includegraphics[scale=.5,viewport=10 250 1000 800,clip]{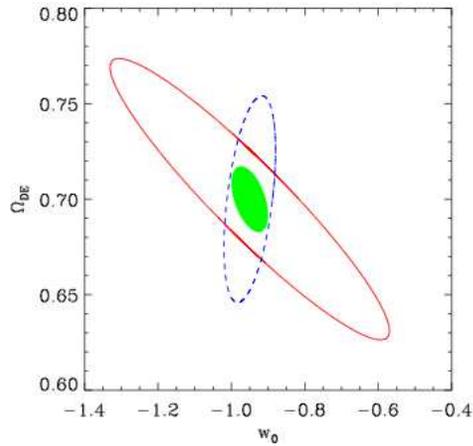}
\caption{Marginalised 68\% CL constraints on the dark energy pa
rameters  expected for the DUNE weak lensing 
(blue), a full sky BAO survey (red) and their combination (solid 
green). This figure was derived using the Fisher matrix routines 
of iCosmo. Figure From Refregier et al 2008.}
\label{fig:icosmo}       
\end{figure}

\subsection{Computing Fisher matrices}

The simplest, brute force approach to compute a Fisher\index{Fisher matrix} matrix is as follows:  write down  the likelihood\index{likelihood} for the data given the model. Instead of the data  values (which are not known) use the theory prediction for a fiducial model. This will add a constant term to the  log likelihood which does not depend on cosmology. In the covariance matrix include expected experimental errors. Then take derivatives with respect to the parameters as indicated in Eq. \ref{eq:fisher}.  
 
In the case where the data are Gaussianly distributed  it is possible to compute explicitly  and analytically  the Fisher matrix\index{Fisher Matrix}, in a much more elegant way than above.
\begin{equation}
F_{ij}=\frac{1}{2}Tr\left[ C^{-1}C_{,i}C^{-1}C_{,j}+C^{-1}M_{ij}\right]
\label{eq:fisherelegant}
\end{equation}
where
$M_{ij}=y_{,i}y^T_{,j}+y_{,j}y^T_{,i}$ and $,i$ denotes derivative wit respect to the parameter $\theta_i$.
This is extremely useful: you need to know the covariance matrix (which may  depend on the model and need not to be diagonal) and you need to have a fiducial model $y$ which you know how it depends on the parameter $\theta$. Then the Fisher matrix give you the expected (forecasted) errors. 
Priors or  forecasts results  from other experiments can be easily included by simply adding their Fisher before performing the matrix inversion to obtain the marginal errors.  This is illustrated in Fig.\ref{fig:icosmo},  from \cite{icosmopaper} and produced using the icosmo (http://www.icosmo.org/) software.

 Before we  finish this section let us spell out the following prescription.
 
  Imagine you have compute a large Fisher\index{Fisher matrix} matrix, varying all  parameters  $\Omega_k$, $w_0$, neutrino mass $m_{\nu}$, number of neutrino species $N_{\nu}$, running of the spectral index $\alpha$ etc. Now you want to compute constraints for a standard flat LCDM model. Simply ignore row and columns corresponding to the parameters that you want to keep fixed at the fiducial value before inverting the matrix.
 
 Imagine now that you have a 6 parameters Fisher matrix (say  $H_0$, $\Omega_m$,$\tau$, $\Omega_{\Lambda}$, $n$, $\Omega_b$, $\sigma_8$),   and want to produce  2D plots for the confidence regions for  parameters 2 and 4, say,  marginalized over all other  (1,3,5,6) parameters.
 Invert $F_{ij}$. Take the sub-matrix made by rows and columns corresponding to the parameters of interest (2 and 4 in this case) and invert back this sub matrix.
 
 The resulting matrix, letÕs call it ${\cal Q}$, describes a Gaussian 2D likelihood surface in the parameters 2 and 4 or, in other words,  the chisquare surface for parameters 2,4 - marginalized over all other parameters- can be described by the equation 
 \begin{equation}
 \widetilde{\chi}^2= \sum_{ij}(\theta_i-\theta_i^{fid.}){\cal Q}_{ij}(\theta_j-\theta_j^{fid.})\,.
 \end{equation}

From this equation, getting the errors corresponds to finding the quadratic equation 
solution $\widetilde{\chi}^2=\Delta \chi^2$. For correspondence between $\Delta \chi^2$ and confidence region see the earlier discussion. If you want to make plots, the equation for the elliptical boundary for the joint  confidence region in the sub-space of parameters of interest is: $\Delta = \delta \theta {\cal Q}^{-1}\delta \theta$.  

\section{Example of  Fisher approach applications}
\label{sec:examples}
Here we are going to consider two cases of application of Fisher\index{Fisher matrix} forecasts that are extensively used in the literature.  This section  assumes that the reader is familiar with basic CMB  and large-scale structure concepts, such as: power spectra, error on power spectra, cosmic variance,  window and selection function, instrumental noise and shot noise, redshift space   etc.  Some readers may find this section more technical than the rest of this document: it is possible  to  skip it, and continue reading   from  \S \ref{sec:modeltesting}.

\subsection{CMB}
\label{sec:cmb}
The CMB has become the single dataset that give most constraints on cosmology. As the recently launched Planck satellite will yield the ultimate survey for primary CMB  temperature anisotropies, doing Fisher matrix forecasts of CMB temperature data may very soon   be obsolete. There remain the scope for forecasting constraints from polarization experiemnts, however systematic effects (e.g. foreground subtraction) will likely dominate the statistical errors (see e.g., \cite{verde06} for details). It is still however a good exercise to see how one can set up a Fisher matrix analysis for CMB data.

If  we have a noiseless full sky survey and the initial conditions are Gaussian we can write that the signal in the sky ( i.e. the spherical harmonic transform of the anisotropies)  is gaussianly distributed. 
we can write the signal as 
\begin{equation}
{\bf s}_{\ell}=(a_{\ell}^T, a_{\ell}^E,a_{\ell}^B)
\end{equation}
 where $a_{ell}$ denotes the spherical harmonic coefficients fro Temperature, E and B model polarization.
 The covariance matrix  ${\bf C}_{\ell}$ is then given by
 \begin{equation}
 {\bf C}_{\ell}=
  \left( \begin{array}{ccc}
C_{\ell}^{TT} & C_{\ell}^{TE}  & 0 \\
C_{\ell}^{TE}   & C_{\ell}^{EE}   & 0\\
0 & 0 &C_{\ell}^{BB}   \end{array} \right)
 \end{equation}
 where $C_{\ell}$ denotes the angular CMB power spectrum.
Using Eq. \ref{eq:fisherelegant} and considering that, for rotational invariance,  for every $\ell $ there are $(2 \ell+1)$ modes, it is possible to show that the Fisher matrix for CMB experiements can be rewritten as:
\begin{equation}
F_{ij}^{CMB}=\sum_{XY}\sum_{\ell}\frac{\partial C_{\ell}^X}{\partial\theta_i} \left({\cal C}_{\ell}^{XY}\right)^{-1}\frac{\partial C_{\ell}^Y}{\partial \theta_j}
\label{eq:fisgercmbquick}
\end{equation}

where the matrix ${\cal C}_{\ell}$ which elements are ${\cal C}_{\ell}^{XY}$, where X,Y=TT, TE, EE, BB etc., is given by\footnote{I owe this proof to P. Adshead}: 
\begin{equation}
{\cal C}_{\ell}=\frac{2}{2 \ell+1}\left( \begin{array}{cccc}
(C_{\ell}^{TT})^2 & (C_{\ell}^{TE})^2 & C_{\ell}^{TT}C_{\ell}^{TE} & 0 \\
(C_{\ell}^{TE})^2   & (C_{\ell}^{EE})^2 &  C_{\ell}^{EE}C_{\ell}^{TE}& 0\\
C_{\ell}^{TT}C_{\ell}^{TE}\,\, & C_{\ell}^{EE}C_{\ell}^{TE}\,\,& 1/2[(C_{\ell}^{TE})^2+C_{\ell}^{TT}C_{\ell}^{EE}] &0\\
0 & 0 &0 &(C_{\ell}^{BB})^2  \end{array} \right)
\end{equation}
Note that this matrix is more complicated that what one would have obtained by assuming a a Gaussian distribution for the $C_{\ell}$ and no correlation between TT TE and EE. Nevertheless Eq. \ref{eq:fisgercmbquick} is simple enough and allows one to quickly compute forecasts from ideal CMB experiments.

In this formalism effects of partial sky coverage and of instrumental noise can be included (approximatively) by the following substitutions:
\begin{equation}
C_{\ell}\longrightarrow C_{\ell}+N_{\ell}  
\end{equation}
where $N_{\ell}$ denotes the effective noise power spectrum. Note that $N_{\ell}$ depends on $\ell$ even for a perfectly white noise because of beam effects.
In addition the partial sky coverage can be accounted for by considering that the number of independent modes decreases with the sky coverage: if $f_{sky}$ denotes the fraction of sky covered by the experiment then 
\begin{equation}
{\cal C}_{\ell}\longrightarrow {\cal C}_{\ell}/f_{sky}
\end{equation}

\subsection{Baryon Acoustic oscillations}
\label{sec:BAO}
Cosmological perturbations in the early universe excite sound waves in the photon-baryon 
fluid. After recombination, these baryon acoustic oscillations (BAO) became frozen into the 
distribution of matter in the Universe imprinting a preferred scale, the sound horizon. This 
defines a standard ruler whose length is the distance sound can travel between the Big Bang 
and recombination. The BAO are directly observed  in the CMB angular power spectrum and 
have been observed in the spatial distribution of galaxies by the 2dF GRS survey and the 
SDSS survey \cite{BAO}. The BAO, observed at different cosmic epochs, act as a powerful 
measurement tool to probe the expansion of the Universe, which in turns is a crucial handle to constrain the nature of dark energy. 
The underlying physics which sets the sound horizon scale ($\sim$150 Mpc comoving) is well 
understood and involves only linear perturbations in the early Universe. The BAO scale is 
measured in surveys of galaxies from the statistics of the three-dimensional galaxy positions. 
Only recently have galaxy surveys such as SDSS grown large enough to allow for this 
detection. The existence of this natural standard measuring rod allows us to probe the 
expansion of the Universe. The angular size of the oscillations in the CMB revealed that the 
Universe is close to flat. Measurement of the change of apparent acoustic scale in a statistical 
distribution of galaxies over a large range of redshift can provide stringent new constraints on 
the nature of dark energy. The acoustic scale depends on the sound speed and the propagation 
time. These depend on the matter to radiation ratio and the baryon-to-photon ratio. CMB 
anisotropy measures these and hence fixes the oscillation scale. A BAO survey measures the 
acoustic scale along and across the line of sight. At each redshift, the measured angular 
(transverse) size of oscillations, $\Delta \theta$, corresponds with the physical size of the sound horizon, 
where the angular diameter distance $D_A$ is an integral over the inverse of the evolving Hubble 
parameter, $H(z)$. $r_{\perp}=(1+z)D_A(z)\delta\theta$. In the radial direction, the BAO directly measure the instantaneous 
expansion rate $H(z)$, through  $r _{\parallel}=(c/H(z)) \Delta z$, where the redshift interval ($\Delta z$) between the peaks is the oscillation scale in the radial direction.
As the true scales $r_{\perp}$ and $r_{\parallel}$ are known (given by $r_s$ the sound horizon at radiation drag, well measured by the CMB)  this is not an Alcock-Paczynsky test but a ``standard ruler'' test.  Note that in this standard ruler test the cosmological feature used as the ruler is not an actual object but a statistical property: a feature in the galaxy correlation function (or power spectrum).
An unprecedented experimental effort is undergoing to obtain galaxy surveys that are deep, larger and accurate enough to trace the BAO feature as a function of redshft. Before these surveys can even  be designed it is crucial to know how well a survey with given characteristic will do. This was illustrated very clearly in \cite{seoEisenstein1}, which we follow closely here. 
We will adopt the Fisher\index{Fisher matrix} matrix approach. To start we need  to compute the statistical error associated to a determination of the galaxy power spectrum $P(k)$. In what follows we will ignore effects of non-linearities  and complicated biasing between galaxies and dark matter: we will assume that galaxies, at least on large scales, trace the linear matter power spectrum in such a way that their power spectrum is directly proportional to the dark matter one: $P(k)=b^2P_{DM}(k)$ where $b$ stands for galaxy bias.
At a given wavevector $k$, the statistical error of the power spectrum is a sum of a cosmic variance term and a shot noise term:
\begin{equation}
\frac{\sigma_P(k)}{P(k)}=\frac{P(k)+1/n}{P(k)}
\label{eq:pkerr}
\end{equation}
Here $n$ denotes the average density of galaxies and $1/N$ is the white noise contribution from the fact that galaxies are assumed to be a Poisson sampling of the underlying distribution. When written in this way this expression assumes that $n$ is constant with position. While in reality this is not true for 
forecasts one assumes that the survey can be divided in shells in redshifts and that the selection function is such that $n$ is constant within a given shell. Since $P(k)$ is also expected to change in redshift   then one should really implicitly assume that there is a $z$ dependence in Eq. \ref{eq:pkerr}. In general $P(k,z)=b(z)^2G^2(z)P_{DM}(k)$ where $G(z)$ denotes the linear growth factor: i.e. the bias is expected to evolve with redshift as well as clustering does, not only because galaxy bias changes with redshift but also because at different redshifts one may be seeing different type of galaxies which may have different bias parameter.  We do not know a priori  the form of $b(z)$ but given a fiducial cosmological model we know $G(z)$. Preliminary observations seem to indicate that the $z$ evolution of $b$ tends to cancel that of $G(z)$, so it is customary to assume that $b(z)G(z)\sim constant$, but we should bear in mind that this is an assumption.

An extra complication arises because galaxy redshft surveys use the redshift as distance indicator, and deviations from the Hubble flow therefore distort the clustering. If the Universe was perfectly uniform and galaxies were test particles these deviations from the Hubble flow would not exist and the survey would not be distorted. But clustering does perturb the Hubble for and thus introduces the so-called redshift-space distortions in the clustering measured by galaxy redshift surveys. Note that redshft-space distortions only affect the line-of-sigth clustering (it is a perturbation to the distances) not the angular clustering.  Since these distortions are created by clustering they carry,  in principle, important cosmological information. To write this dependence explicitly:
\begin{equation}
P(k,\mu,z)=b(z)^2 G(z)^2P_{DM}(k)(1+\beta \mu)^2
\end{equation}
 where $\mu$ denotes the cosine of the angle between the line of sight and the wavevector. $\beta=f/b= d\ln G(z)/d \ln a /b \simeq \Omega_m(z)^{0.6}/b$. In the linear regime, the cosmological information carried by the redshft space distortions  is enclosed in the $f(z)=\beta(z) b(z)$ combination.
 
For finite surveys, $P(k)$ at  nearby wavenumbers  are highly correlated, the correlation length is related to the size of the survey volume: for large volumes the cell size over which modes are correlated is $(2 \pi)^3/V$ where $V$ denotes the comoving survey volume. Only over distances in k-space larger than that modes can be considered independent. If one therefore wants to count over all the modes anyway (for example by transforming discrete sums into integrals in the limit of large volumes) then each $k$ needs to be downweighted, to account the fact that all $k$ are not independent.  In addition one should keep in mind that Fourier modes $\vec{k}$ and $-\vec{k}$ are not independent (the density field is real-valued!), giving an extra factor of 2 in the weighings.
We can thus write the error on a band power centered around $k$,
\begin{equation}
\frac{\sigma_P}{P}=2 \pi \sqrt{\frac{2}{V k^2 \delta k \Delta \mu}}\left(\frac{1+nP}{nP}\right)\,.
\end{equation}

In the spirit of the Fisher approach we now assume that the Likelihood function for the band-powers $P(k)$ is Gaussian thus we can approximate the Fisher matrix by:
\begin{equation}
F_{ij}=\int_{k_{min}}^{k_{max}}\frac{\partial \ln P(\vec{k})}{\partial \theta_i}\frac{\partial \ln P(\vec{k})}{\partial \theta_j} V_{eff}(\vec{k}) \frac{d\vec{k}}{2(2\pi)^3}
\label{eq:fisherbao1}
\end{equation}
 The derivatives should be evaluated at the fiducial model and $V_{eff}$ denotes the effective survey volume given by
 \begin{equation}
 V_{eff}(\vec{k})=V_{eff}(k,\mu)=\int \left[\frac{n(z)P(k,\mu)}{n(z)P(k,\mu)+1}\right]^2dz=\left[\frac{nP(k,\mu)}{nP(k,\mu)+1}\right]^2V
 \end{equation}
where $n=\langle n(z)\rangle$.
Eq.\ref{eq:fisherbao1} can be written explicitely as a function of $k$ and $\mu$ as:
\begin{equation}
F_{ij}=\int_{-1}^{1}\int_{k_{min}}^{k_{max}}=\frac{\partial \ln P(k,\mu)}{\partial \theta_i}\frac{\partial \ln P(k,\mu)}{\partial \theta_j} V_{eff}(k,\mu) \frac{ k^2 dk d\mu}{2(2\pi)^2}\,.
\label{eq:fisherbao2}
\end{equation}

In writing this equation we have assumed that over the entire survey extension the line-of-sight direction does not change: in other words we made the flat sky approximation. For forecasts this encloses all the statistical information anyway, but for actual data-analysis  application the flat sky approximation may not hold.  In this equation $k_{min}$ is set by the survey volume: for future surveys where the survey volume is large enough to sample the first BAO wiggle the exact value of  $k_{min}$ does not matter, however, recall that for surveys of typical size $L$ (where $L\sim V^{1/3}$), the largest scale probed by the survey will be corresponding to $k= 2\pi/L$. Keeping in mind that the first BAO wiggle happens at $\sim  150 $ Mpc the survey size needs to be $L \gg 150$ Mpc for $k_{min}$ to be unimportant and for the "large volume approximation" made here to hold.
As anticipated above, one may want to sub-divide the survey in independent redshift shells,  compute the Fisher matrix for each shell and then combine the constraints. In this case  $L$ will be set by the smallest dimension of the volume (typically the width of the shell) so one needs to make sure that the width of the shell still guarantees   a large volume and large $L$.
$k_{max}$ denotes the maximum wavevector to use. One could for example impose a sharp cut to delimit the range of validity of linear theory.  In \cite{seoEisenstein2} this is  improved  as we will see below.

Before we do that, let us note that there are two ways to interpret the parameters $\theta_{ij}$ in  Eq. (\ref{eq:fisherbao2}). One could simply assume a cosmological model, say for example a flat quintessence model where  the equation of state parameter $w(z)$ is parameterized by $w(z)=w(0)+w_a(1-a)$  and take derivatives of $P(k,\mu)$ with respect to these parameters.  Alternatively,  one could simply use as parameters the quantities $H(z_i)$ and $D_A(z_i)$, where $z_i$ denote the survey redshift bins. These are the quantities that govern the BAO location and are more general: they  allow one not to choose a particular dark energy model until the very end. Then one must also consider   the cosmological parameters  that govern the $P(k)$ shape $\Omega_m h^2$, $\Omega_b h^2$  and $n_s$. Of course one can also consider $G(z_i)$ as free parameters and constrain these either through the overall $P(k)$ amplitude (although one would have to assume that $b(z)$ is known, which is dicey) or  through the determination of $G(z)$ and $\beta(z)$. The safest and most conservative  approach  however is to ignore any possible information coming from  $G(z)$, $\beta(z)$ or $n_s$  and to only try to constrain expansion history parameters. 

The piece of information still needed is how the expansion history information is extracted from $P(k,\mu)$. 
When one  converts {\bf ra}, {\bf dec}  and redshifts into distances  and positions of galaxies of a redshift survey, one assumes a particular reference cosmology.  If the reference cosmology differs from the true underlying cosmology, the inferred distances will be  wrong and so the observed power spectrum will be distorted:
\begin{equation}
P(k_{\perp},k_{\parallel})=\frac{D_a(z)^2_{ref}H(z)_{true}}{D_A(z)_{true}^2 H(z)_{ref}}P_{true}(k_{\perp},k_{\parallel}).
\label{eq:bao3}
\end{equation}
Note that  since distances are affected by the choice of cosmology and k vectors are: $k_{ref,\parallel}=H(z)_{ref}/H(z)_{true}k_{true,\parallel}$ and $k_{ref,\perp}= D_A(z)_{true}/D_A(z)_{ref}k_{true,\perp}$.
Note that therefore in Eq.\ref{eq:bao3} we can write:
\begin{equation}
P_{true}(k_{\perp},k_{\parallel},z)=b(z)^2\left(1+\beta(z) \frac{k^2_{true,\parallel}}{k_{true,\perp}^2+k_{true,\parallel}^2}\right)^2\left[\frac{G)(z)}{G(z_o)}\right]^2P_{DM}(k,z_o)
\end{equation}
where $z_o$ is some reference redshift where to normalize $P(k)$ typical choices can be the CMB redshift or redshift $z=0$.
Not that from thises equations it should be clear that what the BAO actually measure directly is $H(z)r_s$ and $D_A/r_s$ where $r_s$ is the BAO scale, the advantage is that $r_s$ is determined exquisitely from the CMB. 

\begin{figure}[t]
\includegraphics[scale=.42]{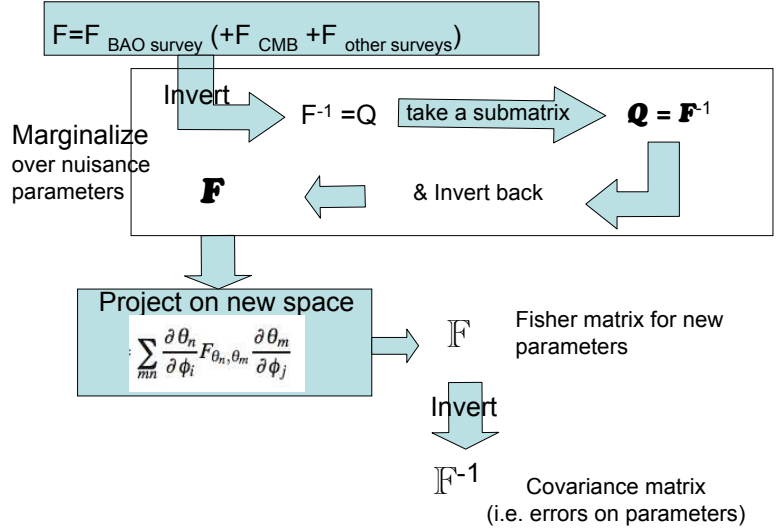}
\caption{Steps to implement once the Fisher matrix of Eq.\ref{eq:fisherbao2} has been computed to obtain error on dark energy parameters.}
\label{fig:projectfisher}       
\end{figure}

How would then one convert these constraints on those on a model parameter? Clearly, one then projects the resulting Fisher matrix on the dark energy parameters space. In general if you have  a set of parameters $\theta_i$ with respect to which the Fisher matrix has been computed, but you would like to have the Fisher matrix for a different set of parameters $\phi_i$ where  the $\theta_i$ are functions of the $\phi_i$, the operation to implement is:
\begin{equation}
F_{\phi_i,\phi_j}=\sum_{mn}\frac{\partial \theta_n}{\partial\phi_i} F_{\theta_n,\theta_m}\frac{\partial \theta_m}{\partial\phi_j}
\end{equation}

the full procedure for the BAO survey case  is illustrated in Fig. \ref{fig:projectfisher}. The slight complication is that one starts off with a Fisher matrix (for the original parameter set  $\theta_i$) where some parameters are nuisance and need to be marginalized over, so some matrix inversions are needed. 

So far non-linearities have been just ignored. It is however possible to include then at some level  in this description. \cite{seoEisenstein2} proceed by introducing a distribution of  gaussianly distributed  random displacements  parallel or perpendicular to the line of sight coming from non-linear growth (in all directions) and from non-linear redshift space distortions (only in the radial direction). The publicly available code that implements all this (and more) is at:\\ http://cmb.as.arizona.edu/~eisenste/acousticpeak/bao$_-$forecast.html.
In order to use the code keep in mind that \cite{seoEisenstein2} model the the effect of non-linearities is  to convolve the galaxy distribution with a redshift dependent and $\mu$ dependent smoothing kernel. The effect on the power spectrum is to multiply   $P(k)$ by $\exp[-k^2 \Sigma(k,\mu)/2]$, where $\Sigma(k,\mu)= \Sigma^2_{\perp}-\mu^2(\Sigma^2_{\parallel}-\Sigma^2_{\perp})$. As a consequence the integrand of  the Fisher matrix expression of Eq. (\ref{eq:fisherbao2}) is multiplied by
\begin{equation}
\exp[-k^2 \Sigma^2_{\perp}-k^2\mu^2(\Sigma^2_{\parallel}-\Sigma^2_{\perp}) ]
\end{equation}
where, to be conservative,  the exponential factor has been taken outside the derivatives, which is equivalent to marginalize over the parameters $\Sigma_{\parallel}$ and $\Sigma_{\perp}$ with large uncertainties.

Note that $\Sigma_{\parallel}$ and $\Sigma_{\perp}$ depend on redshift and on the chosen normalization for $P_{DM}(k)$. In particular: 
\begin{eqnarray}
\Sigma_{\perp}(z)&=&\Sigma_0 G(z)/G(z_0) \\
\Sigma_{\parallel}(z)&=&\Sigma_0 G(z)/G(z_0)(1+f(z))\\
\Sigma_0&\propto& \sigma_8\,.
\end{eqnarray}
 If in your convention $z_0=0$ then $\Sigma_0(z=0)=8.6 h^{-1}\sigma_{8,DM}(z=0)/0.8$.

\begin{figure}[t]
\includegraphics[scale=.33]{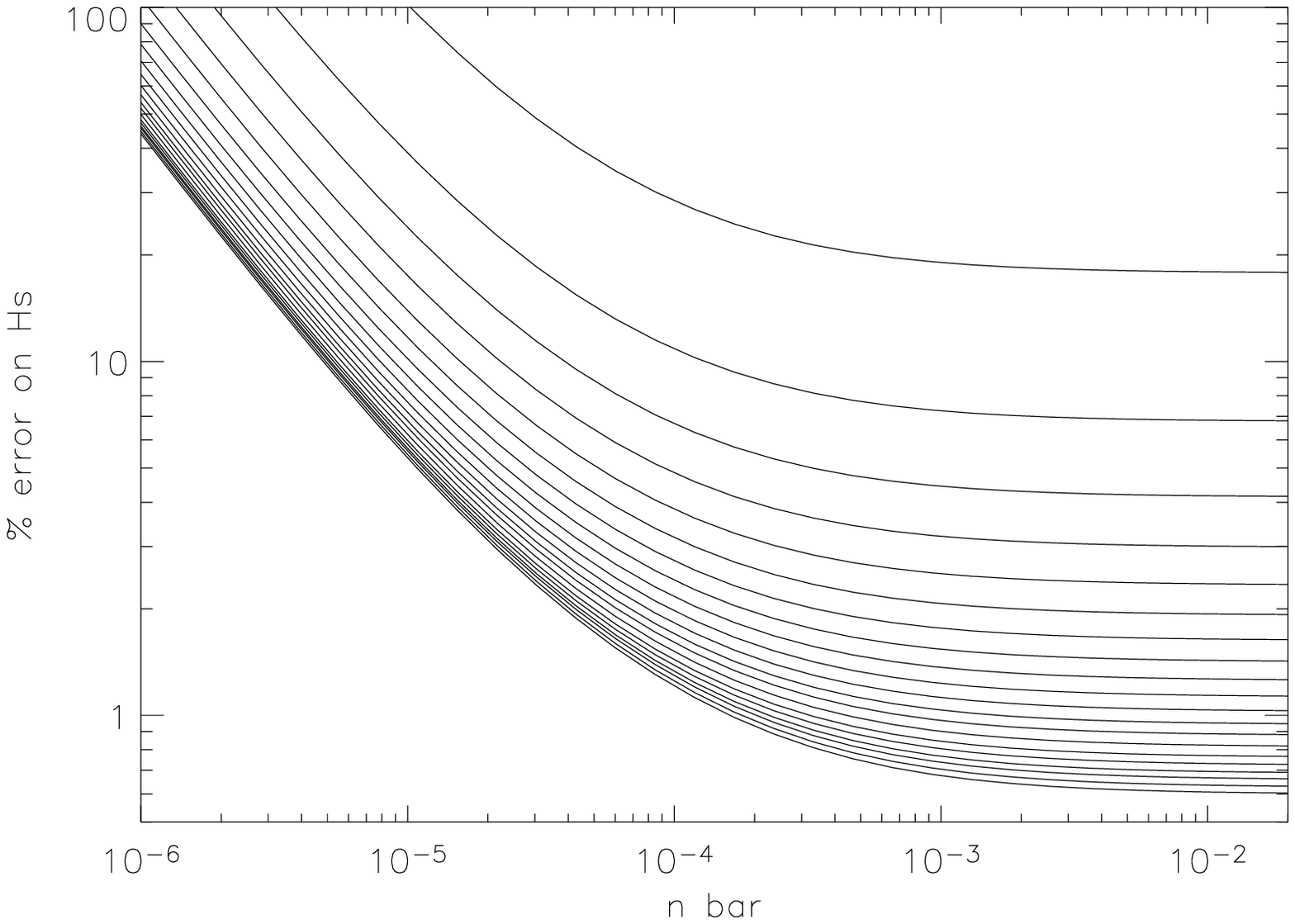}
\includegraphics[scale=.33]{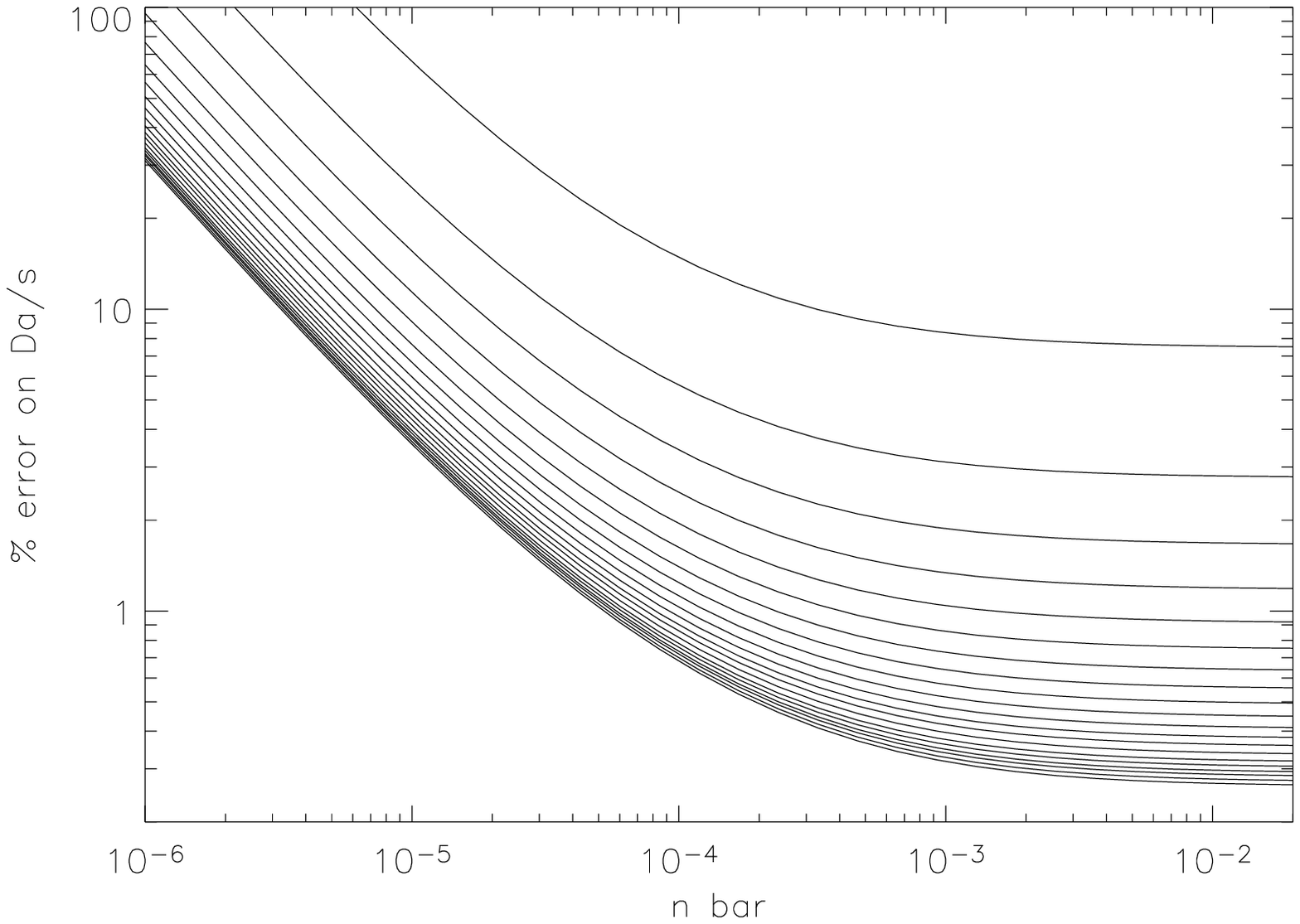}
\caption{Percent error on $H(z)r_s$ and $D_a/r_s$ as a function of the galaxy number density of a BAO survey. This figure assumes full sky coverage $f_{sky}=1$ (errors will scale like $1/\sqrt{f_{sky}}$) and redshift range from $z=0$ to $z=2$ in bins of $\Delta z=0.1$.
\label{fig:np} }      
\end{figure}

As an example of an application of this approach for survey design, 
it may be interesting to ask the question of  what is the optimal galaxy number density for a given survey. 
Taking redshifts is expensive and for a given telescope time allocated, only a certain number of redshifts can be observed. Thus is it better to survey more volume but have a  low number density or survey a smaller volume with higher number density? You can try to address this issue using the available code. For a cross check, figure \ref{fig:np} show what you should obtain. Here we have assumed $\sigma_8=0.8$ at $z=0$, $b(z=0)=1.5$ and we have assumed that $G(z)b(z)=constant$. 
To interpret this figure note that  with the chosen normalizations, $P(k)$ in real space at the BAO scale $k\sim 0.15$ h/Mpc is 6241(Mpc/h)$^3$, boosted up by large scale redshift space distortions to roughly $10^{4}$(Mpc/h)$^3$ so $n=10^{-4}$ corresponds to $nP(k=0.15)=1$. Note that the ``knee'' in this figure is therefore around $nP=1$. This is where this ``magic number" of reaching $nP\sim >1$ in a survey comes from.  Of course, there are other considerations that would tend to yield an optimal $nP$ bigger than unity and of order of few.

\section{Model testing}
\label{sec:modeltesting}
\begin{figure}[t]
\includegraphics[scale=.34]{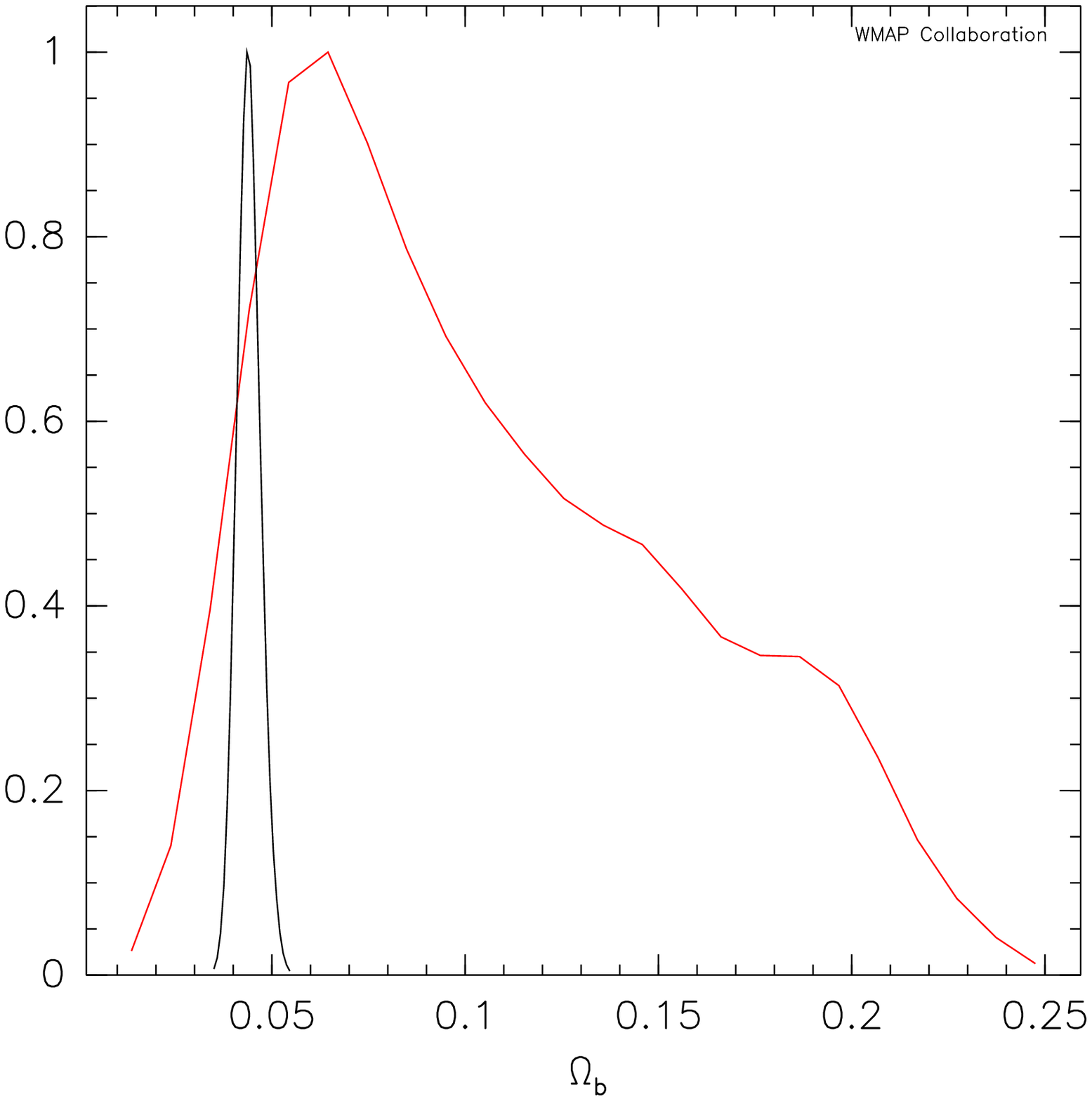}
\includegraphics[scale=.34]{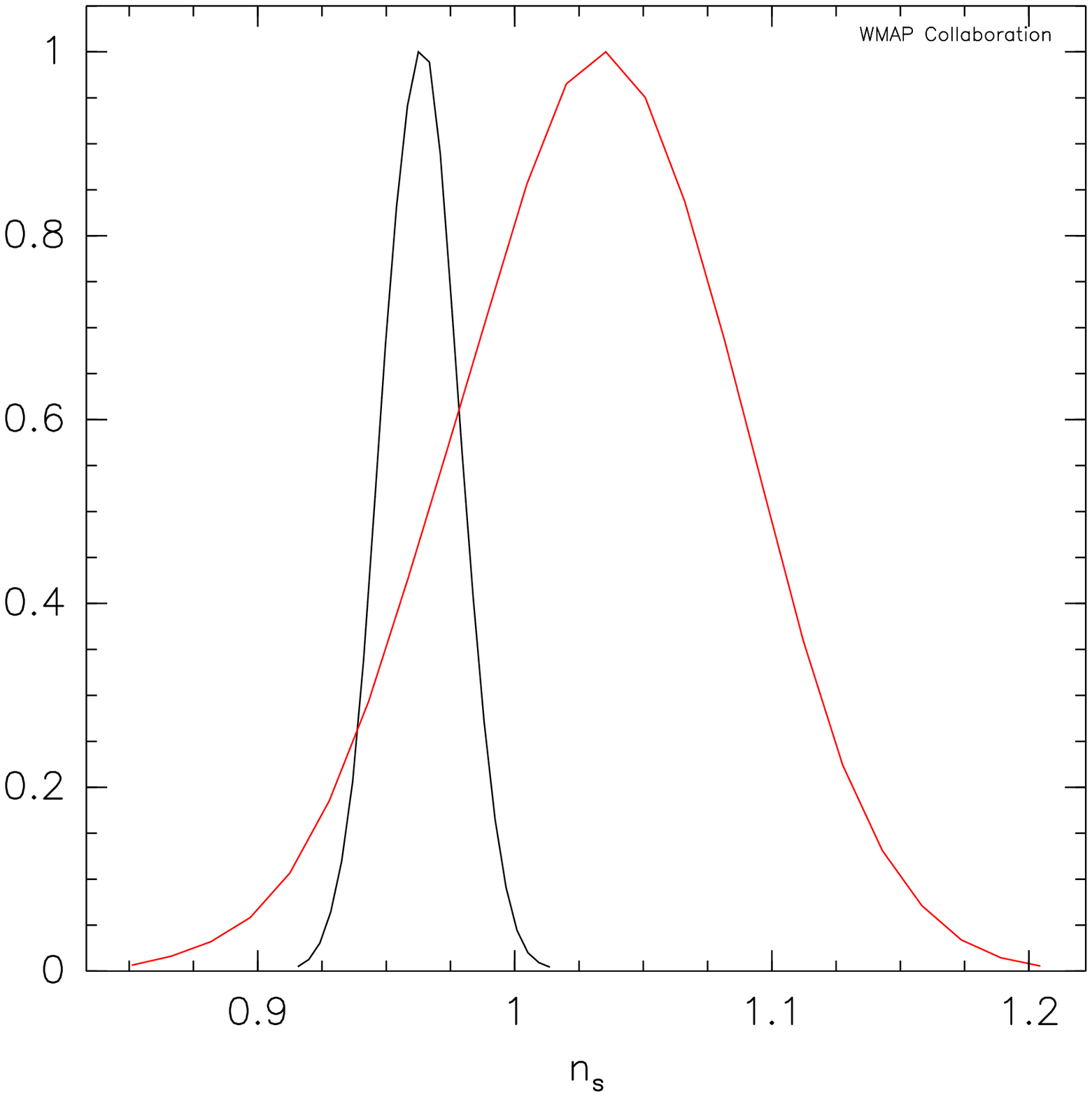}
\caption{Effect of the choice of the cosmological model in the recovered values for the parameters.  Here we used WMAP5 data only: in both panels the black line is for a standard flat LCDM model.  In the left panel   we show the posterior for $\Omega_b$, the red line is for a non-flat LCDM model. In the right panel we show the posterior for $n_s$: the red line is for a LCDM model where the primordial power spectrum is not a perfect power law but is allowed to have some ``curvature" also called ``running" of the spectral index.  Figure courtesy of LAMBDA \cite{LAMBDA}.}
\label{fig:models}       
\end{figure}

So far we have assumed a cosmological model characterized by a given set of cosmological parameters and used statistical tools to determine the best fit for these parameters and confidence intervals.  However the best fit  parameters and confidence intervals  depends on the underlying model i.e. what set of parameters are allowed to vary. For example  the estimated value for the density parameter of baryonic matter $\Omega_b$ changes depending wether in a LCDM model  the universe is assumed flat or not (Fig.\ref{fig:models} right panel) or the recovered value for the spectral slope of the primordial power spectrum changed depending if the primordial power spectrum is assumed to be a power law or is allowed to have some  ``curvature" or ``running" (Fig.\ref{fig:models} left panel).
It would be useful to be able to allow the data to determine which combination of parameters gives the preferred fit to the data: this is the problem of {\it model selection}. Here we start by following \cite{Liddle04} which is a clear introduction to the application of this subject in cosmology.  
Model selection  relies on the so-called ``information criteria"  and the goal is to make an objective comparison of different {\it models} which may have a different number of parameters. The models considered in the example of Fig, \ref{fig:models} are ``nested" as one model (the LCDM one) is  completely specified by a  sub-set of the parameters of the other (more general) model.  In cosmology one is almost always concerned with nested models. 

Typically the introduction of extra parameters  will yield an improved fit yo the data set, so a simple comparison of the maximum likelihood value will always favor the model with more parameters, regardless of wether the extra parameters are relevant.  
There are several different approaches often used in the literature.
The simplest is the likelihood ratio test \cite{kendall3} see \S \ref{sec:5}. Consider the quantity $2 \ln [{\cal L}_{simple}/{\cal L}_{complex}]$ where ${\cal L}_{simple}$ denotes the maximum likelihood for the model with less parameters and ${\cal L}_{complex}]$ the maximum likelihood for the other model.  This quantity is approximately chisquare distributed and thus the considerations of sec \ref{sec:3} can be applied.  

The Akaike information criterion (AIC) \cite{AIC} is defined as:
$AIC=-2\ln {\cal L}+2 k$ where ${\cal L}$ denotes the maximum likelihood for the model and $k$ the number of parameters of the model. The best model is the one that minimizes AIC.

The Bayesian information criterion  (BIC)\cite{BIC} is defined as:
$BIC=-2\ln {\cal L} +k \ln N$ where $N$ is the number of data points used in the fit. 

It should be clear that all these approaches tend to downweight  the improvement in the  likelihood value  for the more complex model with a penalty that depends on how complex is the model. Each of these approaches has its pros and cons and there is no silver bullet.

However it is possible to place  model selection on firm statistical grounds within the Bayesian\index{Bayesian}approach by using the {\it Bayesian factor} which is the Bayesian  evidence\index{Bayesian Evidence} ratio (i.e. the ratio of probabilities of the 
data given the two models).

Recalling the Bayes theorem (Eq.\ref{Eq:bayesth}) we can write: ${\cal P}(D)=\sum_i {\cal P}(D|M_i){\cal P}(M_i)$  where $i$ runs over the models  $M$ we are considering. Then the Bayesian Evidence is
\begin{equation}
{\cal P}(D|M_i)=\int d\theta {\cal P}(D|\theta,M_i){\cal P}(\theta | M_i)
\end{equation}
where $ {\cal P}(D|\theta,M_i)$ is the likelihood. 
Given two models ($i$ and $j$), the Bayes factor is 
\begin{equation}
B_{ij}=\frac{{\cal P}(D|M_i)}{{\cal P}(D|M_j)}.
\end{equation}
A large $B_{ij}$ denotes preference for model $i$. 
 In general this requires complex numerical calculations, but for the simple case of Gaussian likelihoods it can be expressed analytically. The details can be found e.g. in \cite{HeavensBayes} and references therein. For a didactical introduction see also \cite{heavensstats}.

\section{Monte Carlo methods}
\label{sec:montecarlo}
With the recent increase in computing power, in cosmology we resort to the application of  Monte Carlo methods   ever more often. There are two main applications of Monte Carlo methods: Monte Carlo error estimations and Markov Chains Monte Carlo. Here I will concentrate on the first as there are several  basics and details explanations of the second  (see e.g. \cite{LVstats} and references therein).

LetÕs go back to the issue of parameter estimation and error calculation. Here is the 
conceptual interpretation of what it means that en experiment measures some parameters 
(say cosmological parameters). There is some underlying true set of parameters $\theta_{true}$ that 
are only known to Mother Nature but not to the experimenter. There true parameters 
are statistically realized in the observable universe and random measurement errors are 
then included when the observable universe gets measured. This ÒrealizationÓ gives the 
measured data 
$ D_0$ . Only $ D_0$ is accessible to the observer (you). Then you go and do what 
you have to do to estimate the parameters and their errors (chi-square, likelihood, etc.) 
and get $\theta_0$. Note that 
$ D_0$ is not a unique realization of the true model given by $\theta_{true}$:
there could be infinitely many other realizations as hypothetical data sets, which could have 
been the measured one: 
$ D_2, D_2,  D_3...$ each of them with a slightly different fitted parameters $\theta_1$, 
$\theta_2$ ..... $\theta_0$ 
is one parameter set drawn from this distribution. The hypotetical ensamble 
of universes described by $\theta_i$ is called ensamble, and one expects that the expectation value $\langle \theta_i \rangle=\theta_{true}$. If we knew the distribution of $\theta_i-\theta_{true}$ we would know everything we need 
about the uncertainties in our measurement $\theta_0$ . The goal is to infer the distribution of $\theta_i-\theta_{true}$ without knowing $\theta_{true}$. 
HereÕs what we do: we say that hopefully $\theta_{0}$ is not too wrong and we consider a fictitious 
world where $\theta_{0}$ was the true one. So it would not be such a big mistake to take the 
probability distribution of  $\theta_i-\theta_{0}$ to be that of $\theta_{i}-\theta_{true}$ . In many cases we know how to 
simulate  $\theta_i-\theta_{0}$  and so we can simulate many synthetic realization of ``worldsÓ where $\theta_0$ is 
the true underlying model. Then mimic the observation process of these fictitious Universes 
replicating all the observational errors and effects and from each of these fictitious universe 
estimate the parameters. Simulate enough of them and from $\theta^S_i-\theta_{0}$  )where $S$ stands for ``synthetic" or ``simulated") you will be able to 
map the desired multi-dimensional probability distribution. 
With the advent of fast computers this technique has become increasingly widespread. 
As long as you believe you know the underlying distribution and that you believe you can 
mimic the observation replicating all the observational effects this technique is extremely 
powerful and, I would say, indispensable.  This is especially crucial when complicated effects such as instrumental and or systematic effects can be simulated but not described analytically by a model.

\section{Conclusions}
\label{sec:conlcusions}
I have given a brief overview of statistical techniques that are frequently used in the cosmological literature. I have presented several examples often from the literature to put these techniques into context. 
This is not an exhaustive list nor a rigorous treatment, but a starter kit  to ``get you started".  
As more and more sophisticated  statistical techniques are used to make the most of  the data, one should always remember that they need to be implemented and used correctly:
\begin{itemize}
\item data gathering is an expensive and hard task: statistical techniques make possible  to make the most of the data
\item always beware of systematic effects 
\item  an incorrect treatment of the data will give non-sensical results 
\item   there will always be  things that are beyond the statistical power of a given  data set
\end{itemize}
Remember: ``Treat your data with respect!"

\section{Some useful references}
\label{sec:usefulrefs}
There are many good and rigorous statistics books out there. In particular Kendall's advanced theory of statistics  made of three volumes: 
\begin{itemize}
\item Distribution theory (Stuart \& Ort 1994)\cite{kendall1}
\item Classical Inference (Stuart \& Ort 1991) \cite{kendall2} and
\item Bayesian Inference (O'Hagan 1994)\cite{kendall3}.
\end{itemize}
For astronomical and cosmological applications in many cases one may need a practical manual rather than a rigorous textbook. Although it is important to note that a practical manual is no substitute for a rigorous introduction to the subject.
\begin{itemize}
\item Practical statistics for Astronomers, by Wall \& Jenkins, (2003) is a must have \cite{Wall}.
\item Numerical Recipes is also an indispensable ``bible": Press et al (1992)\cite{numrec}
\end{itemize}
 It also provides a guide to  the numerical implementation of the ``recipes" discussed.
Complementary information to what presented here  can be found in 
\begin{itemize}
\item{} Verde, in XIX Canary Island Winter School "The Cosmic Microwave Background: from Quantum fluctuations to the present Universe" \cite{LVstats}, In the form of lecture notes,  and 
\item{}   Martinez, Saar, "Statistics of the galaxy distribution" \cite{Martinez}, with a slant on Large scale structure and Data Analyais in Cosmology, Martinez, Saar, Martinez-Gonzalez, Pons-Porteria, Lacture Notes in Physics 665, Springer, 2009

\end{itemize}

\begin{acknowledgement}
I would like to thank the organizers of the Transregio Tonale winter School 2008 for a truly enjoyable week.  LV is supported by FP7- PEOPLE-2002IRG4-4-IRG\#202182 and MICINN grant AYA2008-03531.
I acknowledge  the use of the Legacy Archive for Microwave Background Data Analysis (LAMBDA). Support for LAMBDA is provided by the NASA Office of Space Science.

\end{acknowledgement}
%
%
%
%

\end{document}